\documentclass[final,1p,times]{elsarticle}
\usepackage{amssymb}
\usepackage{amsmath}
\usepackage{amsthm}
\makeatletter
\@ifundefined{textcolor}{}
{
 \definecolor{BLACK}{gray}{0}
 \definecolor{WHITE}{gray}{1}
 \definecolor{RED}{rgb}{1,0,0}
 \definecolor{GREEN}{rgb}{0,1,0}
 \definecolor{BLUE}{rgb}{0,0,1}
 \definecolor{CYAN}{cmyk}{1,0,0,0}
 \definecolor{MAGENTA}{cmyk}{0,1,0,0}
 \definecolor{YELLOW}{cmyk}{0,0,1,0}
}

\@ifundefined{date}{}{\date{15-Jan-2022}}
\@ifundefined{textcolor}{}
{
 \definecolor{BLACK}{gray}{0}
 \definecolor{WHITE}{gray}{1}
 \definecolor{RED}{rgb}{1,0,0}
 \definecolor{GREEN}{rgb}{0,1,0}
 \definecolor{BLUE}{rgb}{0,0,1}
 \definecolor{CYAN}{cmyk}{1,0,0,0}
\definecolor{MAGENTA}{cmyk}{0,1,0,0}
 \definecolor{YELLOW}{cmyk}{0,0,1,0}
}

\@ifundefined{date}{}
{\date{Sep-2022}}
\makeatother

\begin{document}
\begin{frontmatter}
\title{Geodesic Deviation in S\'{a}ez--Ballester Theory}

\author[1]{S. M. M. Rasouli}
%\email{mrasouli@ubi.pt}
\author[2]{M. Sakellariadou}
%\email{mairi.sakellariadou@kcl.ac.uk}
\author[1]{Paulo Vargas Moniz}
%\email{pmoniz@ubi.pt}

\address[1]{Departamento de F\'{i}sica,
Centro de Matem\'{a}tica e Aplica\c{c}\~{o}es (CMA-UBI),
Universidade da Beira Interior,
 Rua Marqu\^{e}s d'Avila
e Bolama, 6200-001 Covilh\~{a}, Portugal.}

%\affiliation{Department of Physics, Qazvin Branch, Islamic Azad University, Qazvin, Iran.}
\address[2]{Theoretical Particle Physics and Cosmology Group, Physics Department,
King's College London,\\ University of London, Strand, London WC2R 2LS, UK.}

\begin{abstract}
We study the geodesic deviation (GD) equation in
 a generalized version of the S\'{a}ez--Ballester (SB) theory
 in arbitrary dimensions. We first establish
 a general formalism and
 then restrict to particular cases, where (i) the matter-energy
 distribution is that of a perfect fluid, and (ii)
 the spacetime geometry is described by a vanishing Weyl tensor.
 Furthermore, we consider the spatially flat FLRW
 universe as the background geometry.
 Based on this setup, we compute the GD equation as well as the
  convergence condition associated with fundamental
observers and past directed null vector fields.
 Moreover, we extend that framework
and extract the corresponding geodesic deviation
in the \emph{modified} S\'{a}ez--Ballester theory (MSBT),
where the energy-momentum tensor and
potential emerge strictly from the geometry of the extra dimensions.
In order to examine our herein GD equations, we consider two
novel cosmological models within the SB framework. Moreover, we
discuss a few quintessential models and a suitable
   phantom dark energy scenario within the mentioned SB and MSBT frameworks.
   Noticing that our herein cosmological models
can suitably include
 the present time of our Universe,
 we solve the GD equations analytically and/or numerically. By employing the correct energy
 conditions plus recent observational data, we consistently depict the behavior of
  the deviation vector $\eta(z)$ and the observer area distance $r_0(z)$ for
 our models. Concerning the Hubble constant
 problem, we specifically focus on the observational data reported by the Planck
collaboration and the SH0ES collaboration to
depict $\eta(z)$ and $r_0(z)$ for our herein phantom model.
 Subsequently, we contrast our results with those associated
 with the $\Lambda$CDM model. We argue that the MSBT can be considered as a fitting
 candidate for a proper description of the late evolution of the universe.
 \end{abstract}

\medskip

\end{frontmatter}
%\maketitle

\section{Introduction}
\label{int}
\indent
The literature referring to scalar--tensor
theories applied to investigate problems in cosmology
 is vast,
 e.g.,~\cite{Faraoni.book,Q19,K19} and references therein.
S\'aez and Ballester may have been inspired by scalar--tensor theories and formulated a theory that is completely different with scalar--tensor theorieis\footnote{In Refs \cite{RM18,RPSM20,R22}, once the
S\'aez--Ballester theory has been introduced, it was erroneously included in the class of scalar--tensor theories. It is important to emphasize that such statements has not affected the formulation and consequences of those works and we ensure that they are fully correct.} in its construction and motivation \cite{SB85}. More concretely, in the SB theory the scalar field with a rather specific non--canonical kinetic term is added to the Einstein--Hilbert
action. There was no scalar potential but a Lagrangian associated with ordinary matter was also considered. With this modification,
the SB theory suggested a way to overcome the `missing matter problem' in
cosmology~\cite{CCLRD11}, which was actually the motivation
with which the SB framework
was originally proposed~\cite{SB85}.
 Since then, the SB theory has  been  appraised within classical cosmology in, e.g.,~\cite{P87,SA91,SS03,MSM07,NSR12,RKN12,Y13,RPR15}, whereas
in~\cite{SSL10} a quantization with the Wheeler--DeWitt equation was reported.

Notwithstanding the significant
references associated with S\'aez--Ballester (SB) theory  in particular, regarding cosmological applications, it seems, with
respect to the scalar--tensor theories, that it
has been much less investigated.
Although it is worth noting that SB theory is an attractive area of research due to the recent generalization \cite{RPSM20}.

The above-mentioned reasons have ingrained a robust motivation for our endeavor to explore and extract physical, testable consequences from SB cosmology~\cite{RM18,RPSM20} yet on investigate the geodesic deviation (GD) construction in this paper. It is worthy noting that the GD equation has not been investigated within the SB theory.

    The geodesic deviation (GD) equation
is a pertinent tool
to study properties of curved spacetimes~\cite{S34,P56,EE97,P57}.
It has been extensively investigated within different gravitational
theories, by means of various
exact cosmological solutions (see, e.g.,~\cite{GCT11,SHJ16,RBJF09,DMA15,RS21}
and references therein).
Therefore, we establish the GD
formalism associated with the  SB and
MSBT alike~\cite{SB85,RPSM20}. The latter
is a new generalized version of the SB theory
and is established by a dimensional reduction procedure upon the
geometry of extra dimensions.
 The Lagrangian associated with the matter plus
a scalar potential are present, with the number of
dimensions assumed to be arbitrary but where, crucially,
 an effective energy--momentum tensor (EMT) and
a potential are dictated from the geometry, instead of being added
by \textit{ad hoc} assumptions.
As the particular case of MSBT, by considering a five-dimensional manifold
 (empty of ordinary matter), this yields
an effective framework on a four-dimensional hypersurface,
in which the usual right hand side of the
field equations is explained solely in terms of the
whole geometry.
This effective framework is called the space--time matter
theory or the induced-matter theory
(IMT)~\cite{PW92,stm99,5Dwesson06,OW97,LRTR97}
     (see also~\cite{RM18,RPSM20,ARB07,DRJ09,RJ10,RFS11,RFM14,R14,RM16,RM19},
as cosmological applications).

 In this context, the main objectives of our paper are as follows:
(i) To formulate  the GD equation in a SB
 theory (either the original or extended settings) in arbitrary dimensions.
(ii) To obtain \textit{new} exact solutions
 in the context of a SB theory.
 This will allow us to apply the GD equation to
 pertinent cosmological case studies. In particular, we will consider the
 quintessential and a phantom dark energy scenarios
 and contrast them  with results found for the $\Lambda$CDM model.
             (iii)
   To demonstrate that, although all
 the mentioned models yield similar behaviors for
 selected  observables, as far as the MSBT setting is concerned, it
 still constitutes  most satisfactorily a fair and realistic
 route to describe the evolution of the late current and late universe.

In the next section, after introducing an
extended version of the SB theory in arbitrary dimensions, we investigate the corresponding GD equation.
We first obtain the GD equation and then
formulate it according to:
(i) a line-element implying a vanishing Weyl tensor,
(ii) a perfect fluid as the matter-energy sector,
and (iii) a Friedmann--Lema\^{\i}tre--Robertson--Walker (FLRW) metric as the background geometry.
 Subsequently, we focus on the GD equation for
fundamental observers and the null vector field past directed.
In \ref{GR}, by assuming a constant scalar
field, we show that the formalism obtained in \ref{Null} reduces
to that associated with GR in the presence
of the cosmological constant in arbitrary dimensions.
In \ref{SB-GDE}, we  obtain  the energy conditions (i.e.,~weak
 energy condition (WEC), null energy condition (NEC), strong
  energy condition (SEC) and dominant energy condition (DEC)) within the context of SB cosmology,
  and then study the GD equation associated with null vector field in the SB framework.
In order to apply the formalism obtained in
 \ref{SetUp}, we extract \textit{new} exact cosmological solutions in the context of the
SB theory in the absence of the ordinary matter.
We show that these solutions can be applied to
describe the accelerating late time epoch.
Moreover, we investigate the GD equation associated
with a phantom dark energy model and compare the
results with the corresponding ones associated with the $\Lambda$CDM model.
In  \ref{MSBT}, we review the MSBT framework,
and then explore with similar detail as well the GD equation in this context.
In  \ref{Concl}, we present our conclusions.

\section{GD equation in the context of the generalized SB theory in arbitrary dimensions}
\label{SetUp}
\indent
%In this section,
Let us retrieve the GD equation associated with the
generalized SB theory in a $D$-dimensional spacetime and in the presence of a general scalar potential.

The action associated with a $D$-dimensional SB theory, in analogy with the
corresponding four-dimensional case~\cite{SB85}, in the
presence of a scalar potential $V(\phi)$, can be written as
\begin{eqnarray}\label{induced-action}
 {\cal S}^{^{(D)}}=\int d^{^{\,D}}\!x \sqrt{-g}\,
 \Big[R^{^{(D)}}-{\cal W}\phi^n\, g^{\alpha\beta}\,({\nabla}_\alpha\phi)({\nabla}_\beta\phi)
 -V(\phi)+\,
L\!^{^{(D)}}_{_{\rm matt}}\Big],
\end{eqnarray}
where $g$ and $R^{^{(D)}}$ stand for the determinant
and Ricci scalar associated with the
$D$-dimensional metric $g_{\alpha\beta}$, respectively.
 %The
 Greek indices run from zero to $D-1$ and $\nabla$ denotes
the covariant derivative on the $D$-dimensional spacetime.
Throughout this work we use %the
units where $8\pi G=1=c$
(where $G$ and $c$ are the Newton gravitational constant and the
speed of light, respectively). Moreover, $\phi$ is a dimensionless scalar field
(which is hereafter designated as the SB scalar field),
 ${\cal W}$ and $n$ are two dimensionless parameters of the model.
The Lagrangian associated with the ordinary matter fields
is denoted by $L\!^{^{(D)}}_{_{\rm matt}}$, which is independent of the SB scalar field.

The equations of motion obtained
from the action \eqref{induced-action} are:
\begin{eqnarray}\label{BD-Eq-DD}
G_{\mu\nu}^{^{(D)}}=
 T_{\mu\nu}^{^{(D)}} +
{\cal W}\phi^n\left[({\nabla}_\mu\phi)({\nabla}_\nu\phi)-
\frac{1}{2}g_{\mu\nu}({\nabla}_\alpha\phi)
({\nabla}^\alpha\phi)\right]
-\frac{1}{2}g_{\mu\nu}V(\phi)
\end{eqnarray}
and
\begin{eqnarray}\label{D2-phi}
2\phi^n{\nabla}^2\phi+n\phi^{n-1}({\nabla}_\alpha\phi)({\nabla}^\alpha\phi)
-\frac{V_{,\phi}}{\cal W}=0,
\end{eqnarray}
where $\nabla^2\equiv \nabla_a{\nabla}^a$ and $V_{,\phi}\equiv \delta V/\delta\phi$.
Here $T^{^{(D)}}_{\mu\nu}$ and $G^{^{(D)}}_{\mu\nu}$ denote
the EMT (associated with the ordinary matter) and
the Einstein tensor, respectively. Moreover, one can easily show
\begin{eqnarray}\label{cons.law}
{\nabla}_\mu T^{^{(D){}\mu\nu}}=0.
\end{eqnarray}
%
%In what follows,
Let us first obtain the general expression for the GD equation corresponding to the generalized SB
theory (in the presence of the scalar potential) in $D$ dimensions
without choosing a line-element or any constraints on the EMT. Subsequently, we assume that the
 matter %in the universe
 is a perfect fluid, which simplifies our expressions.
 We will then consider the spatially flat $D$-dimensional FLRW line-element as
 the background metric, and investigate the GD equation
 for different cases associated with the generalized SB framework.

Let $\gamma_1$ and $\gamma_2$ be two neighboring geodesic curves,
both parameterized by $\zeta$.
Consider $\textbf{v}$ and $\pmb{\eta}$ as the tangent vector to the
curves and the connecting vector (which connects two points of $\gamma_1$ and $\gamma_2$
with the same value of the parameter $\zeta$), respectively.
%Indeed,
The GD of the curves is measured by $\pmb{\eta}$.
Assuming $\textbf{v}$ and $\pmb{\eta}$ as the coordinate basis vectors of a
coordinate system, %then
we have $[\pmb{\eta},\textbf{v}]=0$.
Subsequently, using $\nabla_{\textbf{v}} \textbf{v}=0$
(%because
the curves have been assumed as geodesics)
and the antisymmetry property for the Riemann tensor \textbf{R}, it
is straightforward to show that
\begin{equation}\label{GDE-gen-0}
 \nabla_{\textbf{v}}\nabla_{\textbf{v}}
 \pmb{\eta}+\textbf{R}(\pmb{\eta},\textbf{v})\textbf{v}=0,
\end{equation}
which is the GD equation.
%As the component form of the
%equations will be useful in this paper, hence let us
%rewrite equation \eqref{GDE-gen}
Equivalently, it can be rewritten as \cite{GH.book}
\begin{eqnarray}\label{GDE-gen}
\left(\frac{{ d}^2\pmb{\eta}}{{ d}\zeta^2}\right)^\alpha=
-R^\alpha_{\,\,\beta\gamma\delta}v^\beta \eta^\gamma v^\delta,
\end{eqnarray}
which implies that the measurements of the
GD can determine completely the Riemann tensor.
%In this paper, we will
In our analysis, we assume that the tangent vector
field $v^\alpha\equiv\frac{dx^\alpha(\zeta)}{d\zeta}$ is normalized as
\begin{eqnarray}\label{t-vector2}
v_\alpha v^\alpha=\varepsilon,
\end{eqnarray}
where $\varepsilon=-1,0,1$ correspond to the timelike, null and
spacelike geodesics, respectively. Moreover, as mentioned, $\pmb{\eta}$
commutes with $\pmb{v}$, i.e., $\eta_\alpha v^\alpha={\rm constant}$.
Therefore, without loss of generality, we %will further
take
\begin{eqnarray}\label{t-vector3}
\eta_\alpha v^\alpha=0.
\end{eqnarray}
In $D$ dimensions (for $D\geq3$), in the component form, we
have\footnote{From now on, we remove the upper index
 $(D)$ from the quantities.} \cite{Inverno.book}:
\begin{eqnarray}\nonumber
R_{\alpha\beta\gamma\delta}\!\!&=&\!\!C_{\alpha\beta\gamma\delta}
+\frac{1}{D-2}\left(g_{\alpha\gamma}R_{\delta\beta}-g_{\alpha\delta}
R_{\gamma\beta}+g_{\beta\delta}R_{\gamma\alpha}-g_{\beta\gamma}R_{\delta\alpha}\right)\\
\label{Rie-ten}
\!\!&+&\!\!\frac{R}{(D-1)(D-2)}\left(g_{\alpha\gamma}g_{\delta\beta}
-g_{\alpha\delta}g_{\gamma\beta}\right),
\end{eqnarray}
where $C_{\alpha\beta\gamma\delta}$ is the Weyl
tensor.
In order to obtain a more useful expression associated with the right hand side (r.h.s.) of
equation \eqref{GDE-gen}, we
%will proceed according to the following steps:
do the following:
(i) We only consider the background metrics whose Weyl
tensor vanishes. (ii) We raise the first index of the Riemann tensor
and then contract it with $v^\beta \eta^\gamma v^\delta$.
(iii) In order to simplify, we compute explicitly the Ricci tensor and the Ricci curvature
 scalar. More concretely, equation \eqref{BD-Eq-DD} yields
\begin{eqnarray}\label{gen-Ricc-scalar}
R=-\frac{2\,T}{(D-2)}+
{\cal W}\phi^n(\nabla_\alpha\phi)(\nabla^\alpha\phi)
+\left(\frac{D}{D-2}\right) V(\phi).
\end{eqnarray}
Replacing $R$ from \eqref{gen-Ricc-scalar} to \eqref{BD-Eq-DD} gives
\begin{eqnarray}\label{gen-Ricc-tensor}
R_{\mu\nu}=T_{\mu\nu}-\left(\frac{T}{D-2}\right)g_{\mu\nu}+
{\cal W}\phi^n(\nabla_\mu\phi)(\nabla_\nu\phi)
+\left(\frac{1}{D-2}\right)g_{\mu\nu}V(\phi).
\end{eqnarray}
(iv) Substituting the Ricci tensor and Ricci scalar from
relations \eqref{gen-Ricc-scalar} and \eqref{gen-Ricc-tensor} into
the expression obtained from step (ii), we get
%\begin{widetext}
 \begin{eqnarray}\label{Rie-ten-2}
R^\lambda_{\,\,\beta\gamma\delta}v^\beta \eta^\gamma v^\delta &=&
\frac{1}{(D-2)}\Bigg\{\delta^\lambda_{\gamma}T_{\delta\beta}-\delta^\lambda_\delta
T_{\gamma\beta}+g_{\beta\delta}T^\lambda_{\gamma}-g_{\beta\gamma}T^\lambda_{\delta} \\\nonumber
\\\nonumber
&-&\frac{2}{(D-1)}\Bigg[T+\frac{\cal W}{2}\phi^n(\nabla_\alpha\phi)(\nabla^\alpha\phi)
-\frac{V(\phi)}{2}\Bigg]\Big(\delta^\lambda_{\gamma}g_{\delta\beta}
-\delta^\lambda_\delta g_{\gamma\beta}\Big)\\\nonumber
\\\nonumber
\!\!&+&\!\!{\cal W}\phi^n
\Big[\delta^\lambda_{\gamma}(\nabla_\delta\phi)(\nabla_\beta\phi)
-\delta^\lambda_\delta (\nabla_\gamma\phi)(\nabla_\beta\phi)
+g_{\beta\delta}(\nabla_\gamma\phi)(\nabla^\lambda\phi)
-g_{\beta\gamma}(\nabla_\delta\phi)(\nabla^\lambda\phi)\Big]\Bigg\}v^\beta \eta^\gamma v^\delta.
\end{eqnarray}
%\end{widetext}
(v) We restrict ourselves to the special case where the EMT is taken as perfect fluid:
\begin{eqnarray}\label{perfect}
T_{\mu\nu}=\left(\rho+p\right)u_\mu u_\nu+pg_{\mu\nu},
\end{eqnarray}
where $\rho$ and $p$ denote the energy density and pressure of the fluid.
The trace of \eqref{perfect} reads
\begin{eqnarray}\label{perfect-trace}
T=-\rho+\left(D-1\right)p,
\end{eqnarray}
where we have used $u^\mu u_\mu=-1$.
Therefore, substituting $T_{\mu\nu}$ and $T$ from
relations \eqref{perfect} and \eqref{perfect-trace} into \eqref{Rie-ten-2}
as well as considering $E=-v_\alpha u^\alpha$, $\eta_\alpha u^\alpha=0$, we obtain
%\begin{widetext}
\begin{eqnarray}\nonumber
R^\lambda_{\,\,\beta\gamma\delta}v^\beta \eta^\gamma v^\delta
\!\!&=&\!\!\frac{1}{(D-2)}\Bigg\{ \left(\rho+p\right)E^2+\frac{\varepsilon}{(D-1)}\Big[2\rho-{\cal W}\phi^n (\nabla_\alpha\phi)(\nabla^\alpha\phi)+V(\phi)\Big]\Bigg\}\eta^\lambda
\\\nonumber
\\\nonumber
\!\!&+&\!\!\frac{{\cal W}\phi^n}{(D-2)}
\Bigg\{\Big[\delta^\lambda_{\gamma}(\nabla_\delta\phi)(\nabla_\beta\phi)
-\delta^\lambda_\delta (\nabla_\gamma\phi)(\nabla_\beta\phi)\\
\!\!&+&\!\!g_{\beta\delta}(\nabla_\gamma\phi)(\nabla^\lambda\phi)
-g_{\beta\gamma}(\nabla_\delta\phi)(\nabla^\lambda\phi)\Big]
v^\beta v^\delta\Bigg\}\eta^\gamma,
\label{GDE-gen-0}
\end{eqnarray}
%\end{widetext}
where we have also used relations \eqref{t-vector2} and \eqref{t-vector3}.

It is worth mentioning that all the above obtained
equations are not only valid for the FLRW metric but also for all
metrics whose Weyl tensor vanishes;
although, we have restricted ourselves to the perfect fluid assumption.
In the next subsection, we focus on
a spatially flat FLRW metric as the background in $D$ dimensions.

%%%%%%%%%%%%%%%%%%%%%%%%%%%%%%%%%%%%%%%%%%%%%%
\subsection{GD equation in the SB theory with a FLRW background}

%In this subsection, we first obtain the GDE for the SB framework
In the particular case where the background metric is the spatially flat FLRW,
the $D$-dimensional spacetime is
\begin{equation}\label{ohanlon metric-2}
ds^{2}=-dt^{2}+a^{2}(t)\left(\frac{dr^2}{1-kr^2}+r^2d\Omega_{_{D-2}}^2\right),
\end{equation}
where $k=-1,0,1$, the scale factor is defined by
$a(t)$ and $d\Omega_{_{D-2}}^2=d\theta_1^2+sin^2\theta_1d\theta_2^2+...+sin^2\theta_1
...sin^2\theta_{D-3}d\theta_{D-2}^2$ for $D\geq3$. Due to the spacetime symmetries,
 the components of the metric as well as the SB scalar field depend only
on the cosmic time. Moreover, let us
%confine our attention to
concentrate on the $D$-dimensional flat spacetime for which $k=0$.
Therefore, equation \eqref{GDE-gen-0} reduces to
\begin{eqnarray}\label{FT-SB}
R^\lambda_{\,\,\beta\gamma\delta}v^\beta \eta^\gamma v^\delta
=\Bigg\{E^2\left(\frac{\rho_{_{\rm eff}}+p_{_{\rm eff}}}{D-2}\right)
+2\varepsilon \left[\frac{\rho_{_{\rm eff}}}{(D-1)(D-2)}\right]\Bigg\}\eta^\lambda.
\end{eqnarray}
In equation \eqref{FT-SB}, we defined the effective energy density and effective pressure as
\begin{eqnarray}
\rho_{_{\rm eff}}\equiv \rho+\rho_{\phi},\hspace{10mm}p_{_{\rm eff}}\equiv  p+p_{\phi},
\label{eff.ro-p}
\end{eqnarray}
where
\begin{eqnarray}
\label{rho-phi-gen}
\rho_\phi\!\!&\equiv\!\!&\frac{1}{2}\left[{\cal W}\phi^n\dot{\phi}^2+V(\phi)\right],
\\\nonumber\\
p_\phi\!\!&\equiv\!\!&\frac{1}{2}\left[{\cal W}\phi^n\dot{\phi}^2-V(\phi)\right].
\label{p-phi-gen}
\end{eqnarray}
Indeed, the force term given by equation \eqref{FT-SB} is the
generalized version of the one obtained in \cite{EE97, P56}.

Consequently, by substituting the force term \eqref{FT-SB} into
equation \eqref{GDE-gen}, the GD equation associated with the SB framework
with the FLRW background in $D$ dimensions is% obtained:
\begin{eqnarray}\label{GDE-gen-2}
\frac{{d}^2\eta^\lambda}{{d}\zeta^2}
=-\Bigg\{E^2\left(\frac{\rho_{_{\rm eff}}+p_{_{\rm eff}}}{D-2}\right)
+2\varepsilon \left[\frac{\rho_{_{\rm eff}}}{(D-1)(D-2)}\right]\Bigg\}\eta^\lambda,
\end{eqnarray}
which is the generalized version of the Pirani equation \cite{P56,EE97}. Equation \eqref{GDE-gen-2} implies
that the spatial orientation of the connecting vector is
not included in the GD equation. However, if
%at the start of our computations,
we had not restricted
ourselves to the isotropic symmetry, then the GD equation would have included not only the
magnitude of the connecting vector along the geodesic
 but also its directional change; see for instance, \cite{CCT10}.

\subsubsection{GD equation for a fundamental observer}
%\subsection{GDE for a fundamental observer}
\label{Fundamental}

In this case, $v^\alpha$ and the affine parameter $\zeta$ can be replaced by the
$D$-velocity of the fluid $u^\alpha$ and the proper time $t$, respectively.
Moreover, letting the vector fields be normalized as $E=1$, and considering
temporal geodesics, i.e., $\varepsilon=-1$, equation \eqref{FT-SB} reduces to
\begin{eqnarray}
R^\lambda_{\,\,\beta\gamma\delta}u^\beta \eta^\gamma u^\delta=
\left[\frac{ (D-3)\rho_{_{\rm eff}}+(D-1)p_{_{\rm eff}}}{(D-1)(D-2)}\right]\eta^\lambda.
\label{fun-0}
\end{eqnarray}
Assuming the connecting vector to be $\eta^\lambda=\vartheta e^\lambda$
(where the basis $e^\lambda$ is %parallelly
propagated parallel to %along
the $D$-velocity), then %due to the isotropy, we have
isotropy implies
\begin{eqnarray}
\frac{de^\lambda}{dt}=0,
\label{fun-1}
\end{eqnarray}
which leads to obtain
\begin{eqnarray}
\frac{d^2\eta^\lambda}{dt^2}=\frac{d^2\vartheta}{dt^2}e^\lambda.
\label{fun-2}
\end{eqnarray}
Consequently, the GD equation for this case is written as
\begin{eqnarray}
\frac{d^2{\vartheta}}{dt^2}=-\left[\frac{(D-3)\rho_{_{\rm eff}}+(D-1)p_{_{\rm eff}}}{(D-1)(D-2)}\right]{\vartheta},
\label{fun-3}
\end{eqnarray}
which is the Raychaudhuri
equation associated with the SB theory (in $D$ dimensions and in the
presence of a scalar potential) when the universe is described %with
by a spatially flat FLRW metric in $D$-dimensions. (For a recent investigation of Raychaudhuri
equation, see \cite{BMDHU18,CDB21}.)
Equation \eqref{fun-3} can be applied to both comoving matter
as well as non-comoving one, which, for the particular
case where $\phi={\rm constant}$ and $D=4$, it has been investigated
 in \cite{EE97}. Moreover, from equation \eqref{fun-3}, we see that focusing
 condition for all timelike geodesics is given by
 \begin{eqnarray}
\frac{(D-3)\rho_{_{\rm eff}}+(D-1)p_{_{\rm eff}}}{(D-1)(D-2)}>0.
\label{focous.fun-3}
\end{eqnarray}

In this study, let us merely consider the comoving matter where we
set $\vartheta=a(t)$. Therefore, equation \eqref{fun-3} reduces to
\begin{eqnarray}\label{fun-Ray}
\frac{\ddot{a}}{a}=-\left[\frac{(D-3)\rho_{_{\rm eff}}+(D-1)p_{_{\rm eff}}}{(D-1)(D-2)}\right].
\end{eqnarray}
 We should note that the equation \eqref{fun-Ray} can also be deduced from combining
the field equations associated with the spatially flat FLRW
metric in the context of the SB framework (including a scalar potential):
\begin{eqnarray}
\label{fun-Fri-1}
\frac{(D-1)(D-2)}{2}H^2
\!\!&=&\!\!  \rho_{_{\rm eff}},\\\nonumber\\
\label{fun-Fri-2}
(D-2)\frac{\ddot{a}}{a}+\frac{(D-2)(D-3)}{2}H^2
\!\!&=&\!\!  -p_{_{\rm eff}},\\\nonumber
\label{wave-eq}
2\phi^n\ddot{\phi}+2(D-1)H\phi^n\dot{\phi}+n\phi^{n-1}\dot{\phi}^2\!\!&+&\!\!\frac{V_{,\phi}}{\cal{W}}=0,\\
\end{eqnarray}
where $H\equiv \dot{a}/a$ is the Hubble parameter.
Moreover, we have
\begin{eqnarray}
\label{cons}
\dot{\rho}\!\!&+&\!\!(D-1)H(\rho+p)=0,\\
\label{cons-phi}
\dot{\rho}_\phi\!\!&+&\!\!(D-1)H\left(\rho_\phi+p_\phi\right)=0.
\end{eqnarray}
The consistency of two different procedures for obtaining the Raychaudhuri
equation indicates that all equations of herein model are correct.

\subsubsection{GD equation for a past directed null vector field}
\label{Null}

We now extend our calculations for the past directed null vector fields.
In this case, $v^\alpha=k^\alpha$ and $k_\alpha k^\alpha=0$.
Therefore, the expressions associated with the force term
according to equation \eqref{FT-SB} reduces to
\begin{eqnarray}
R^\lambda_{\,\,\beta\gamma\delta}v^\beta \eta^\gamma v^\delta
=E^2\left(\frac{\rho_{_{\rm eff}}+p_{_{\rm eff}}}{D-2}\right)
\eta^\lambda,
\label{Null-SB}
\end{eqnarray}
which can be considered as the Ricci focusing in our herein SB framework.
%(in the rest of this subsection, will present more discussion concerning it).
Using a parallelly propagated and aligned basis, i.e. admitting
$\frac{{\cal D}e^\lambda}{{\cal D}\zeta}=k^\alpha \nabla_\alpha e^\lambda=0$,
and setting $\eta^\lambda=\eta e^\lambda$, $e_\lambda e^\lambda=1$,
$e_\lambda u^\lambda=e_\lambda k^\lambda=0$
 \cite{EE97}, equation \eqref{GDE-gen-2} reduces to
\begin{eqnarray}
\frac{d^2\eta}{d\zeta^2}=-E^2\left(\frac{\rho_{_{\rm eff}}+p_{_{\rm eff}}}{D-2}\right)\eta.
\label{BD-null}
\end{eqnarray}
From equation \eqref{BD-null}, we see that if the condition
$\rho+p+{\cal W}\phi^n\dot{\phi}^2>0$,
is satisfied, all families of past-directed as well as future-directed null
geodesics will experience focusing.
In the particular case where ${\cal W}=1$ and $n=0$ (the well-known cosmological model with a single scalar field
minimally coupled to gravity), the above mentioned inequality reduces to
$\rho+p+\dot{\phi}^2>0$
which is always satisfied for a special ordinary matter whose energy density and pressure are related as $\rho+p=0$.

%\rd{discussions about Ricci focussing}

For our herein general case, it will be useful to transform equation
\eqref{BD-null} to the corresponding expression, which is written in terms of
the redshift parameter $z$. In this regard, we write
\begin{eqnarray}\label{ops1}
\frac{d}{d\zeta}=\frac{dz}{d\zeta}\frac{d}{dz},
\end{eqnarray}
which yields
\begin{eqnarray}\label{ops2}
\frac{d^2}{d\zeta^2}=\left(\frac{d\zeta}{dz}\right)^{-2}
\left[\frac{d^2}{dz^2}-\left(\frac{d\zeta}{dz}\right)^{-1}\frac{d^2\zeta}{dz^2}\frac{d}{dz}\right].
\end{eqnarray}
Concerning the null geodesics, we can write
\begin{eqnarray}
1+z=\frac{a_0}{a}=\frac{E}{E_0},
\label{redshift}
\end{eqnarray}
where $a_0=1$ is the present value of the scale
factor (throughout this paper, the index $0$ denotes the value of the corresponding quantity at present time $t_0$).
Moreover, regarding the past directed case, using $dt/d\zeta=E=E_0(1+z)$
(note that for a past directed geodesic, while $z$ increases, $\zeta$ decreases)
as well as \eqref{redshift}, we obtain
\begin{eqnarray}\label{ops3}
\frac{d\zeta}{dz}=\frac{1}{E_0H(1+z)^2}.
\end{eqnarray}
Then, using
\begin{eqnarray}\label{ops3-1}
\frac{dH}{dz}=\frac{d\zeta}{dz}\frac{dt}{d\zeta}\frac{dH}{dt}=-\frac{\dot{H}}{H(1+z)},
\end{eqnarray}
we can show that
\begin{eqnarray}\label{ops4}
\frac{d^2\zeta}{dz^2}=\frac{1}{E_0H^3(1+z)^3}\left(\frac{\ddot{a}}{a}-3H^2\right).
\end{eqnarray}
Consequently, substituting ${d\zeta}/{dz}$ and ${d^2\zeta}/{dz^2}$ respectively from \eqref{ops3} and \eqref{ops4}
into \eqref{ops2} %and then applying it for $\eta$,
we get
\begin{eqnarray}\nonumber
\frac{d^2 \eta}{d\zeta^2}=E_0^2H^2(1+z)^4
\left\{\frac{d^2\eta}{dz^2}+\left[\frac{3H^2-{\ddot{a}}/{a}}{H^2(1+z)}\right]\frac{d\eta}{dz}\right\}.\\
\label{mm}
\end{eqnarray}
Using %Equalizing the r.h.s. of
\eqref{fun-Ray}, \eqref{BD-null} , \eqref{redshift},
and \eqref{mm}, %and then
%using the Raychaudhuri equation \eqref{fun-Ray},
we finally obtain the
GD equation associated with the null vector fields past directed in terms of $z$:
\begin{eqnarray}\label{ops5}
\frac{d^2\eta}{dz^2}
+\frac{1}{1+z}\left[3+\frac{(D-3)\rho_{_{\rm eff}}+(D-1)p_{_{\rm eff}}}{(D-1)(D-2)H^2}\right]\frac{d\eta}{dz}
+\left[\frac{\rho_{_{\rm eff}}+p_{_{\rm eff}}}{(D-2)(1+z)^2H^2}\right]\eta=0.
\end{eqnarray}
%where again we have used \eqref{redshift}.
%
Using \eqref{fun-Fri-1}, equation \eqref{ops5} can be written as
\begin{eqnarray}\label{ops5-1}
\frac{d^2\eta}{dz^2}+\frac{1}{1+z}
\left[\frac{D+3}{2}+\frac{ p_{_{\rm eff}}}{(D-2)H^2}\right]
\frac{d\eta}{dz}
+\frac{1}{(1+z)^2}\left[\frac{D-1}{2}+\frac{p_{_{\rm eff}}}{(D-2)H^2}\right]\eta=0.
\end{eqnarray}
%By taking the following variables

Defining
 \begin{eqnarray} \label{ops5-1-sol-1}
x\equiv 1+z,\hspace{10mm} \eta\equiv \frac{y}{x}, \hspace{10mm}
 \ell(x)\equiv \frac{D+3}{2}+\frac{p_{_{\rm eff}}}{(D-2)H^2},
\end{eqnarray}
equation \eqref{ops5-1} %transforms to
can be written as
\begin{eqnarray}\label{ops5-1-sol-2}
\frac{d^2y}{dx^2}+\left[\frac{\ell(x)-2}{x}\right]\frac{dy}{dx}=0,
\end{eqnarray}
with a general solution %for equation \eqref{ops5-1-sol-2} is
\begin{eqnarray}\label{ops5-1-sol-3}
\eta(z)=\frac{\eta_1}{1+z}+\frac{\eta_2}{1+z}
\times\int dz
\left\{Exp\left[\int^z dz'\left(\frac{D-1}{2}+\frac{p_{_{\rm eff}}}{(D-2)H^2}\right)\right]\right\},
\end{eqnarray}
where $\eta_1$ and $\eta_2$ are constants of integration.

In Section \ref{MSBT}, we will show that all of our herein %the
calculations remain valid for the MSBT framework, for which the components of the induced EMT as
well as the induced scalar potential are directly obtained from the
corresponding equations %and none else, rather than from employing
without any ad hoc phenomenological assumptions.
\\

%%%%%%%%%%%%%%%%%%%%
\subsection{GD equation in GR}
\label{GR}
%In this subsection,
Here we obtain the GD equation for null vector
fields within a spatially flat FLRW background associated with a GR (in the presence of the
cosmological constant, $\Lambda$) in arbitrary dimensions.

%However,
Let us first obtain the GD equation in the context of the MSBT
framework by assuming that the perfect fluid has contributions
from both dust and radiation:
  \begin{eqnarray}\label{matt-rad-1}
\rho \!&=&\!(D-1)H_0^2(1+z)^{D-1}\Big[\Omega_{m0}+\Omega_{r0}(1+z)\Big],
\\\nonumber\\
p\!&=&\!H_0^2\Omega_{r0}(1+z)^{D},
\label{matt-rad-2}
\end{eqnarray}
where $\Omega_i\equiv \rho_i/[(D-1)H^2]$ stands for the
dimensionless cosmological density parameters; the
indices $m$ and $r$ refer to the matter and radiation, respectively.
Moreover, in the above equations $p=p_r=\rho_r/(D-1)$ for which
the conservation law is assumed to be satisfied identically.
For this case, equation \eqref{fun-Fri-1} can be written as
\begin{eqnarray}\label{fun-Fri-3}
H^2=\frac{2H_0^2(1+z)^{D-1}}{(D-2)}\Big[\Omega_{m0}
+\Omega_{r0}(1+z)\Big]+H_0^2\Omega_{\rm DE},
\end{eqnarray}
where
\begin{eqnarray}
\label{DE-omega}
\Omega_{\rm DE}\equiv \frac{2\,\rho_{\phi}}{(D-1)(D-2)H_0^2}\,.
\end{eqnarray}

Therefore, equation \eqref{ops5} reduces to
\begin{eqnarray}\label{ops6}
\frac{d^2\eta}{dz^2}+{\cal P}\frac{d\eta}{dz}+{\cal Q}\eta=0,
\end{eqnarray}
where ${\cal P}\!\!=\!\!{\cal P} (H,dH/dz,z,D)$ and ${\cal Q}={\cal Q} (H,dH/dz,z,D)$ are given by

%\begin{widetext}
\begin{eqnarray}\nonumber
{\cal P}\!\!&\equiv&\!\!\frac{3}{(1+z)}\\\nonumber
&+&\frac{(D-1)H_0^2(1+z)^{D-1}\left[(D-3)\Omega_{m0}+(D-2)\Omega_{r0}(1+z)\right]
+\left[(D-3)\rho_{\phi}+(D-1)p_{\phi}\right]}
{2(D-1)H_0^2(1+z)^{D}\left[\Omega_{m0}+\Omega_{r0}(1+z)\right]+2(1+z)\rho_{\phi}},\\
\label{P-Par}
\\\nonumber\\
{\cal Q}&\equiv& \frac{(D-1)}{2(1+z)^2}\,\Bigg\{\frac{H_0^2(1+z)^{D-1}\left[(D-1)\Omega_{m0}+D\Omega_{r0}(1+z)\right]
+\left(\rho_{\phi}+p_{\phi}\right)}
{(D-1)H_0^2(1+z)^{D-1}\left[\Omega_{m0}+\Omega_{r0}(1+z)\right]+\rho_{\phi}}\Bigg\}.
\label{Q-Par}
\end{eqnarray}
%\end{widetext}
It should be emphasized that, up to now, we have
not restricted our attention to the GR limit.
More concretely, we have merely assumed that the ordinary
EMT has components as \eqref{matt-rad-1} and \eqref{matt-rad-2}. Namely, equations
\eqref{fun-Fri-3}-\eqref{Q-Par} still correspond to the generalized SB framework.

It is pertinent to note that the GR limit can be retrieved by assuming
$\phi={\rm constant}$ and $V\equiv2\Lambda$. Concretely, relations \eqref{rho-phi-gen}
and \eqref{p-phi-gen} reduce to $\rho_{\phi}=-p_{\phi}\equiv\Lambda$, where $\Lambda$ is a constant.
For this particular case, equation \eqref{GDE-gen-2} then reduces to
\begin{eqnarray}\label{GDE-gen-3}
\frac{{\cal D}^2\eta^\lambda}{{\cal D}\zeta^2}=-R^\lambda_{\,\,\beta\gamma\delta}v^\beta \eta^\gamma v^\delta
=-\Bigg\{E^2\left(\frac{ \rho+p}{D-2}\right)
+2\varepsilon \left[\frac{\rho+\Lambda}{(D-1)(D-2)}\right]\Bigg\}\eta^\lambda,
\end{eqnarray}
which is a generalization of the Pirani equation \cite{S34,P56,EE97}.
Moreover, admitting the conditions of the GR limit, equation \eqref{DE-omega} yields
\begin{eqnarray}\label{omeg-lam}
\Omega_{\rm DE}= \frac{2 \Lambda}{(D-1)(D-2)H_0^2}\equiv \Omega_\Lambda={\rm constant}.
\end{eqnarray}
Finally, from equation \eqref{ops6}, we retrieve the GD equation
for null vector fields in the context GR$+\Lambda$ in arbitrary dimensions as
%\begin{widetext}

\begin{eqnarray}\nonumber
\frac{d^2\eta}{dz^2}&+&\Bigg\{\frac{(1+z)^{D-1}\left[(D+3)\Omega_{m0}+(D+4)\Omega_{r0}(1+z)\right]+2(D-2)\Omega_\Lambda}
{2(1+z)^D\left[\Omega_{m0}+\Omega_{r0}(1+z)\right]+(D-2)(1+z)\Omega_\Lambda}\Bigg\}\,\frac{d\eta}{dz}
\\\nonumber\\
&+&
\Bigg\{\frac{(D-1) \Omega _{m0}+D (1+z) \Omega _{r0}}{2 (1+z)^2
\left[\Omega _{m0}+(1+z) \Omega _{r0}\right]+(D-2) (1+z)^{3-D} \Omega _{\Lambda }}\Bigg\}\,\eta=0, \label{GR-D-CC}
\end{eqnarray}
%\end{widetext}
which is exactly the same equation obtained in \cite{RS21}, as expected.

In this paper, we will study %and then depict
the behavior of
$\eta(z)$ and the observer area distance, $r_0(z)$, whose definition is:
 \begin{eqnarray}\label{mattig-1}
r_0(z)=\sqrt{\Bigl|{}\frac{dA_0(z)}{d\Omega_s}\Bigr|{}}=
\Bigl|{}\frac{\eta(z')\mid_z}{d\eta(z')/d\ell\mid_{z=0}}\Bigr|{},
\end{eqnarray}
where $A_0$ is the area of the object and $\Omega_s$ stands for the solid angle.
Note that to compute $r_0(z)$ we use $d/d\ell=E_0^{-1}(1+z)^{-1}d/d\zeta=H(1+z)d/dz$ and
assume an initial condition as $\eta(z=0)=0$.

Equation \eqref{GR-D-CC} has been investigated in \cite{RS21} for some cases.
%by presenting exact solutions as well as numerical analysis.
For later use, let us study another interesting case.
Substituting $D=4$ and $\Omega_{r0}=0$ in equations \eqref{matt-rad-1}-\eqref{fun-Fri-3} and \eqref{GR-D-CC}, we get
 \begin{eqnarray}\label{matt-1}
\rho(z) \!&=&\!3H_0^2\Omega_{m0}(1+z)^{3},\hspace{10mm}p=0,\\\nonumber\\
\label{matt-2}
H(z)
\!\!&=&\!\!H_0\left[\Omega_{m0}(1+z)^3+\Omega_{\Lambda}\right]^{\frac{1}{2}},\\\nonumber\\
\frac{d^2\eta}{dz^2}&+&\frac{1}{2}\left[\frac{7\Omega _{m0} (1+z)^3
 +4 \Omega _{\Lambda }}{ \Omega _{m0}(1+z)^4 +\Omega _{\Lambda } (z+1) }\right]\,
 \frac{d\eta}{dz}
 +\frac{3}{2}\left[\frac{ \Omega _{m0} (1+z)}{ \Omega _{m0}(1+z)^3+ \Omega _{\Lambda }}\right]\,\eta=0,
 \label{GDE-matt}
\end{eqnarray}
where $\Omega _{\Lambda }$ is given by \eqref{omeg-lam}.
%It is easy to show that
An exact solution for \eqref{GDE-matt} is
\begin{eqnarray} \label{GDE-matt-sol}
 \eta (z)=N \, _2F_1\left(\frac{1}{3},\frac{1}{2};\frac{4}{3};
 -\frac{(z+1)^3 \Omega _{m0}}{\Omega _{\Lambda }}\right)
 +\frac{M }{z+1}\sqrt[3]{\frac{\Omega _{\Lambda }}{\Omega _{m0}}},
\end{eqnarray}
where $N$ and $M$ are the integration constants, which carry the
dimension of $\eta$, and $_2F_1(a,b;c;z)$ is the hypergeometric function.
%which has the series expansion
%\begin{eqnarray}\label{Hyp-geom}
%_2F_1(a,b;c;z)=\sum _{k=0}^{\infty } \frac{a_k b_k z^k}{k! c_k}.
%\end{eqnarray}
Moreover, using the definition \eqref{mattig-1}, one can show that
\begin{eqnarray}\label{GDE-matt-r}
r_0 (z)=\frac{1}{H_0\sqrt{\Omega _{\Lambda }}}
\Bigg[\, _2F_1\left(\frac{1}{3},\frac{1}{2};\frac{4}{3};-\frac{(z+1)^3 \Omega _{m0}}{\Omega _{\Lambda }}\right)
-\frac{\, _2F_1\left(\frac{1}{3},\frac{1}{2};\frac{4}{3};-\frac{\Omega _{m0}}{\Omega _{\Lambda }}\right)}{z+1}\Bigg].
\end{eqnarray}
For this case, using recent observational data, we will plot the
behavior $\eta (z)$ and $r_0 (z)$ in Sections \ref{SB-GDE} and \ref{MSBT}.

\section{GD equation in the SB theory for cosmological models}
\label{SB-GDE}

To apply the GD equation \eqref{ops5}
we will first investigate the energy
conditions associated with the SB theory and then consider some models based on the
cosmological equations obtained in Section \ref{SetUp}.
Assuming that the SB scalar field dominates the
dynamics during accelerating phase, in \ref{new-sol}, we will
obtain {\it new} cosmological exact solutions in the absence of ordinary matter.
%Moreover, in subsection
In \ref{Phantom}, we will investigate the GD equation of a
phantom dark energy model in the context of the SB theory, and compare the
%corresponding
results with %those
the corresponding ones of the
$\Lambda$CDM model.
\\
\\
%\onecolumngrid
\subsection {Energy conditions in the generalized SB theory}
\label{ECs}
Applying the results found in \cite{SW13,MM20}, one can %easily
show that the energy conditions in the generalized SB theory in arbitrary dimensions are
%\begin{widetext}
\begin{eqnarray}\label{NEC}
{\rm NEC:}\,\, \,\,  \rho_{_{\rm eff}}\!\!&+&\!\! p_{_{\rm eff}}\geq0,\\
\label{WEC}
{\rm WEC:} \,\,\,\,   \rho_{_{\rm eff}}\!\!&+&\!\! p_{_{\rm eff}}\geq0 ,\hspace{10mm} {\rm and}  \hspace{10mm}   \rho_{\rm tot}\geq 0,\\
\label{SEC}
{\rm SEC:} \,\,\,\,   \rho_{_{\rm eff}}\!\!&+&\!\! p_{_{\rm eff}}\geq0, \hspace{10mm}
{\rm and} \hspace{10mm}   (D-3)\rho_{_{\rm eff}}+(D-1)p_{_{\rm eff}}\geq0,\\
\label{DEC}
{\rm DEC:}\,\, \,\,   \rho_{_{\rm eff}}\!\!&\pm&\!\! p_{_{\rm eff}}\geq0, \hspace{10mm} {\rm and} \hspace{10mm}   \rho_{_{\rm eff}}\geq 0.
\end{eqnarray}
%\end{widetext}

Substituting $\rho_{_{\rm eff}}$ and $p_{_{\rm eff}}$ from \eqref{eff.ro-p} into the above conditions, we obtain
%\begin{widetext}
\begin{eqnarray}\label{NEC1}
{\rm NEC:}\!\!\!\!\!\!&&\rho+p+{\cal W}\phi^n\dot{\phi}^2\geq0,\\
\label{WEC1}
{\rm WEC:} \!\!\!\!\!\!&&\rho +\frac{1}{2}\left[{\cal W}\phi^n\dot{\phi}^2+V(\phi)\right]\geq 0,\hspace{15mm} {\rm and} \hspace{10mm} \rho+p+{\cal W}\phi^n\dot{\phi}^2\geq0 ,\\
\label{SEC1}
{\rm SEC:}\!\!\!\!\!\!&&(D-3)\rho+(D-1)p-V(\phi)+(D-2){\cal W}\phi^n\dot{\phi}^2
\geq 0,\hspace{2mm} {\rm and} \hspace{2mm} \rho+p+{\cal W}\phi^n\dot{\phi}^2\geq0,\\
\label{DEC1}
{\rm DEC:}\!\!\!\!\!\! &&V(\phi)\geq 0,\hspace{5mm} \rho +\frac{1}{2}\left[{\cal W}\phi^n\dot{\phi}^2+V(\phi)\right]\geq 0,\hspace{8mm} {\rm and} \hspace{7mm} \rho+p+{\cal W}\phi^n\dot{\phi}^2\geq0,
\end{eqnarray}
%\end{widetext}
where we have used \eqref{rho-phi-gen} and \eqref{p-phi-gen}.

\subsection{Cosmological exact solutions in vacuum}
\label{new-sol}

%In the follows,
We will present two new exact solutions in the absence of ordinary matter in
the context of the generalized SB theory.
For a single scalar field in the absence of ordinary matter,
using equations \eqref{fun-Fri-1} and \eqref{cons-phi}, we obtain
% two key equations as
\begin{eqnarray}\label{key-1}
\frac{\dot{a}}{a}=\sqrt{\frac{{\cal W}\phi^n\dot{\phi}^2+V(\phi)}{(D-1)(D-2)}},
\end{eqnarray}
\begin{eqnarray}\label{key-2}
\frac{1}{{\cal W}\phi^n\dot{\phi}^2}\frac{\frac{d}{dt}\left[{\cal W}\phi^n\dot{\phi}^2+V(\phi)\right]}{\sqrt{{\cal W}\phi^n\dot{\phi}^2+V(\phi)}}=-2\sqrt{\frac{D-1}{D-2}}.
\end{eqnarray}

%%%%%%%%%%%%%%%%%%%%%%%%%%%%%%%%%%%%%%%%%%%%%%%%%%%%%%%%%%%%%%%%%%%%%%%%%%%%%%%%%%%%%%%%%%%%%%%%%%%%%%%%%%%%%%%%%%%%%%%%%%%%%%%%%%%%%%%%
%%%%%%%%%%%%%%%%%%%%%%%%%%%%%%%%%%%%%%%%%%%%%%%%%%%%%%%%%%%%%%%%%%%%%%%%%%%%%%%%%%%%%%%%%%%%%%%%%%%%%%%%%%%%%%%%%%%%%%%%%%%%%%%%%%%%%%%

\subsubsection{Solution I}
\label{Sol-I}

Let us assume that the potential energy is a function of $\phi$ and $\dot{\phi}$,
as\footnote{We will see that such an assumption (see also \eqref{V-II-0}) yields well-known potentials, which
leads to a model that could account for the present epoch.
More concretely, the choices \eqref{A-1} and \eqref{V-II-0} give, respectively, \eqref{Ex1-Vphi}
and \eqref{V-phi-II}, which are the generalized versions of
the exponential, power-law and Mexican-hat potentials.}
\begin{eqnarray}\label{A-1}
V(\phi)=\left(\frac{2}{\Gamma}-{\cal W}\right)\phi^n\dot{\phi}^2,
\end{eqnarray}
where $\Gamma$ is a constant.

Substituting the potential from \eqref{A-1} into equation \eqref{key-2} leads to
\begin{eqnarray}\label{AE-1}
\left(\frac{1}{\phi^{\frac{3n}{2}}\dot{\phi}^3}\right)\frac{d}{dt}\left(\phi^n\dot{\phi}^2\right)=
-\kappa{\cal W}\Gamma\sqrt{\frac{2}{\Gamma}\left(\frac{D-1}{D-2}\right)}\equiv A,
\end{eqnarray}
where we used $\sqrt{\phi^n\dot{\phi}^2}=\kappa \phi^{\frac{n}{2}}\dot{\phi}$ with $\kappa=\pm 1$.
It is straightforward to show that a solution of equation \eqref{AE-1} is
\begin{eqnarray}\label{AES-1}
\phi^{\frac{n}{2}}\dot{\phi}=-\frac{2}{At},
\end{eqnarray}
where we have set the integration constant equal to zero.

%From %differential
Equation \eqref{AES-1}, implies
\begin{equation}\label{AES-2}
\phi(t)=\left \{
 \begin{array}{c}
\left[\phi_i^{\frac{n+2}{2}}-\frac{(n+2)}{A} \,\, {\rm ln}
\left(\frac{t}{t_i}\right)\right]^{\frac{2}{n+2}},
 \hspace{8mm} {\rm for}\hspace{5mm} n\neq-2,\\\\
 \phi_i\left(\frac{t}{t_i}\right)^{-\frac{2}{A}},
  \hspace{29mm} {\rm for}\hspace{5mm} n=-2,
 \end{array}\right.
\end{equation}
%where we defined
%\begin{eqnarray}\label{alpha}
%\alpha\equiv\frac{2}{(D-1)\Gamma}
%\end{eqnarray}
where $\phi_i$ is the value of the SB scalar field at $t=t_i$.

From equations \eqref{key-1}, \eqref{A-1} and \eqref{AES-1}, we
get a power-law relation for the scale factor as
\begin{eqnarray}\label{AES-3}
a(t)&=&a_i\left(\frac{t}{t_i}\right)^{\alpha},\\\nonumber
\alpha&\equiv&
-\frac{2\kappa}{A}\sqrt{\frac{2}{(D-1)(D-2)\Gamma}}
=\frac{2}{(D-1){\cal W}\Gamma}\hspace{5mm}\forall n,
\end{eqnarray}
where $a_i$ is the value of the scale factor at $t=t_i$.
Moreover, employing relations \eqref{A-1} and \eqref{AES-2}, the potential can be obtained
in terms of the cosmic time as well as SB scalar field:
\begin{eqnarray}\label{Ex1-Vt}
V(t)=\left[\frac{4(2-{\cal W}\Gamma)}{A^2\Gamma}\right]\frac{1}{t^{2}},\hspace{10mm} \forall n,
\end{eqnarray}
 and
% \begin{widetext}
\begin{equation}\label{Ex1-Vphi}
V(\phi)=\left \{
 \begin{array}{c}
V_i
{\rm Exp}\left[-\left(\frac{2A}{n+2}\right)
\left(\phi^{\frac{n+2}{2}}-\phi_i^{\frac{n+2}{2}}\right)\right],
 \hspace{19mm} {\rm for}\hspace{8mm} n\neq-2,\\\\
 V_i \left(\frac{\phi}{\phi_i}\right)^{A}
  \hspace{50mm} {\rm for}\hspace{8mm} n=-2,
 \end{array}\right.
\end{equation}
%\end{widetext}
where
 \begin{eqnarray}\label{Vi}
 V_i\equiv \left[\frac{4(2-{\cal W}\Gamma)}{A^2\Gamma}\right]\frac{1}{t_i^{2}}, \hspace{10mm} \forall n.
\end{eqnarray}

\begin{figure}
\centering\includegraphics[width=2.6in]{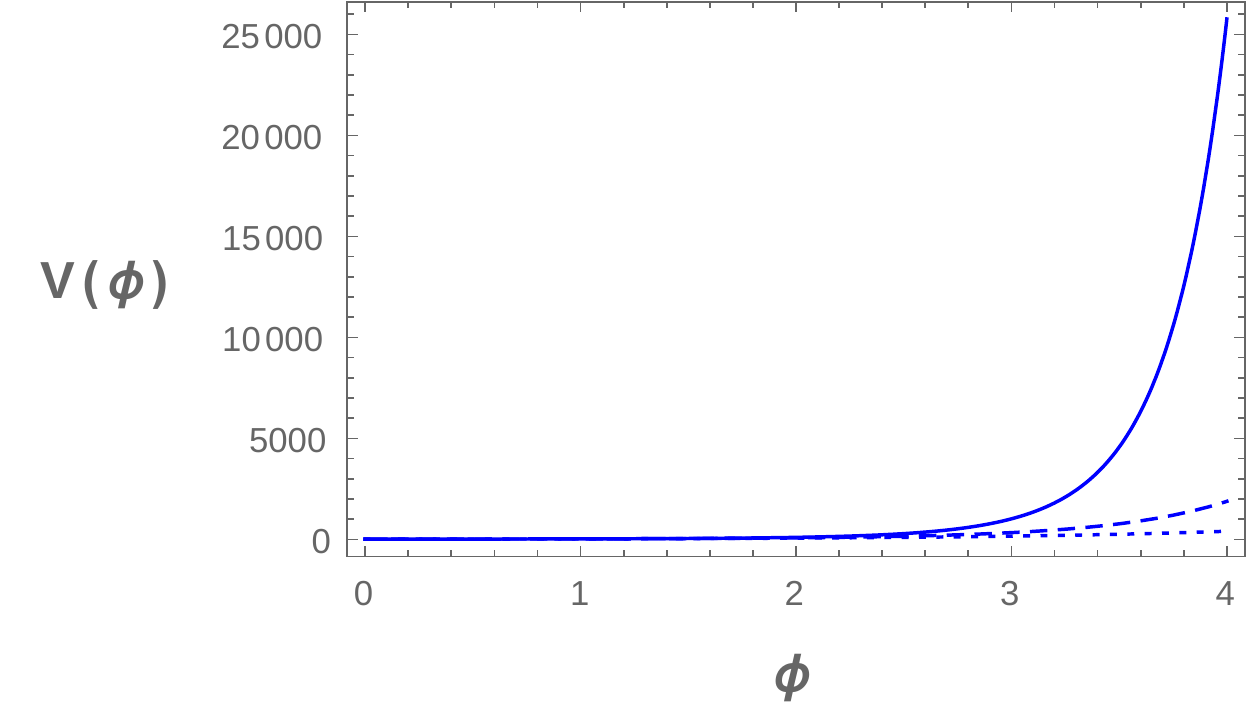}
\centering\includegraphics[width=2.6in]{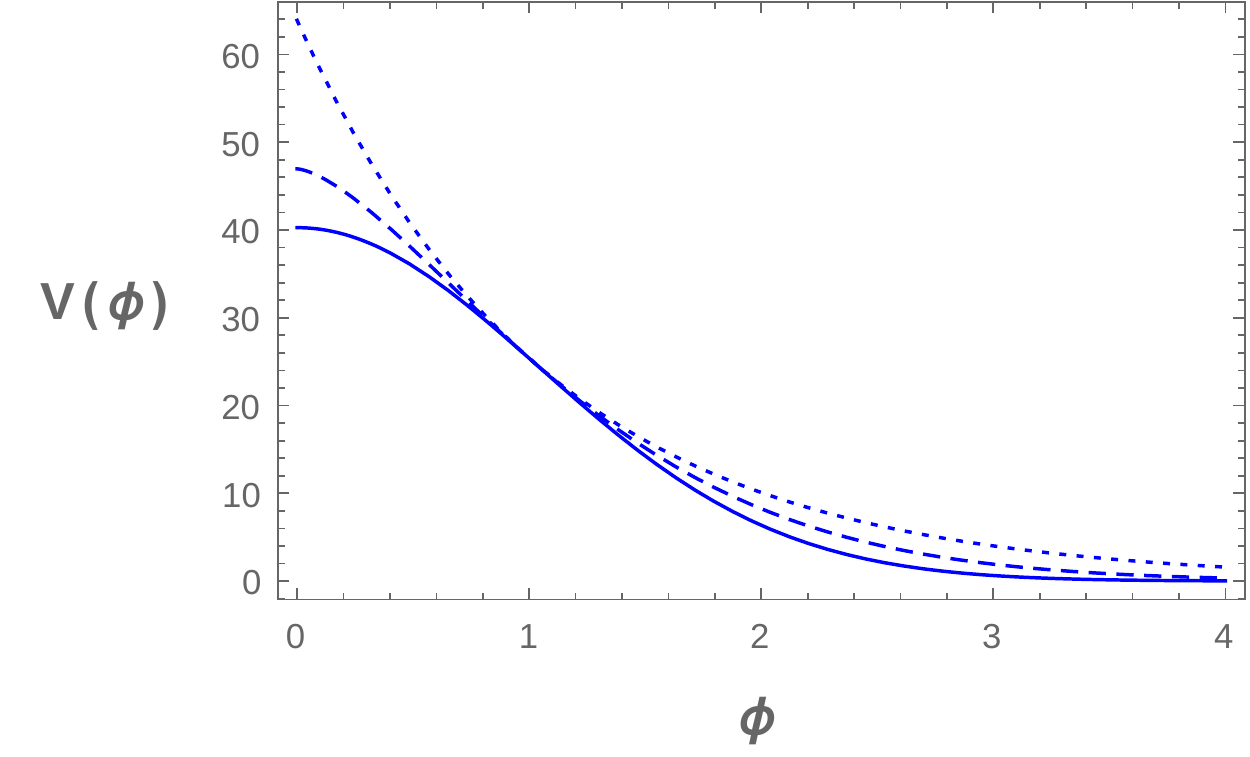}
\caption{{\footnotesize The behavior of $V(\phi)$
associated with the solution I for $\kappa=1$ (the left panel)
and $\kappa=-1$ (the right panel) for different values of $n$: $n=0$ (the dotted curves),
$n=1$ (the dashed curves) and $n=2$ (the solid curves).
We have assumed ${\cal W}=0.95$, $D=4$, $\alpha=2.23$ ($q_0\simeq -0.55$) and $t_i=1=\phi_i$.
%In order to havse all the curves in one panel, we have re-scaled the plots of the right panel.
}}
%\foreignlanguage{english}
{\label{LM-V}}
\end{figure}

Furthermore, for later use, let us also compute the
component of the EMT associated with the scalar field.
From \eqref{rho-phi-gen}, \eqref{p-phi-gen}, \eqref{A-1} and
\eqref{AES-2}, we obtain
\begin{eqnarray}\label{Ex1-ro-phi}
\rho_\phi(t)&=&\left(\frac{4}{A^2\Gamma}\right)\,\frac{1}{t^2},\hspace{20mm} \forall n,\\\nonumber\\
\label{Ex1-p-phi}
p_\phi(t)&=&\left[\frac{4({\cal W}-\Gamma)}{A^2\Gamma}\right]\,\frac{1}{t^2},\hspace{13mm} \forall n,
\end{eqnarray}
which satisfy the conservation law \eqref{cons-phi}, as expected.
%We should mention
Let us note that our solution associated with $n\neq-2$ is a generalized
version of the Lucchin-Mataresse power-law solution \cite{GC07}.
More concretely, for the particular case where $n=0$ and ${\cal W}=1$, the
action \eqref{induced-action} reduces to the
Einstein-Hilbert action including a single scalar field
minimally coupled to gravity, and therefore our herein exact solution
yields the D-dimensional Lucchin-Mataresse power-law solution, as expected.
%It is worthy to depict the behavior of the scalar potential for some values of the present parameters of the model.
As we shown in Fig.\ref{LM-V}, when the SB scalar field grows
the potential increases for $\kappa=1$, while it decreases for $\kappa=-1$.

%In what follows,
Let us investigate the GD equation for the solution I.
Substituting $a(t)$, $\rho_\phi(t)$ and $p_\phi(t)$, respectively
from relations \eqref{AES-3}, \eqref{Ex1-ro-phi} and \eqref{Ex1-p-phi} into
equation \eqref{ops5} and setting $\rho=0=p$, we %easily
get
 \begin{eqnarray}\label{GDE-power}
\frac{d^2\eta}{dz^2}+\left(\frac{{\cal Q}+2}{1+z}\right)\frac{d\eta}{dz}
+\frac{{\cal Q}}{(1+z)^2}\eta=0,
\end{eqnarray}
 where
 \begin{eqnarray}\label{Q}
{\cal Q}\equiv\frac {(D-1){\cal W}\Gamma}{2}.
\end{eqnarray}

It is straightforward to show that %the differential
equation \eqref{GDE-power} yields an exact solution
 \begin{eqnarray} \label{GDE-power-sol}
 \eta (z)= C_1 (z+1)^{-\frac{1}{2}\left [({\cal Q}+1)+\left| {\cal Q}-1\right|\right ]}
 +C_2 (z+1)^{-\frac{1}{2} [({\cal Q}+1)-\left| {\cal Q}-1\right| ]},
 \end{eqnarray}
  where $C_1$ and $C_2$ are the constants of the integration carrying the dimension of $\eta$.

% Employing
Using \eqref{mattig-1} and \eqref{GDE-power-sol} %and \eqref{mattig-1}, it is also easy to show that
the observer area distance is given by
 \begin{eqnarray}\nonumber
 r_0(z)=\frac{(z+1)^{-\frac{1}{2} [({\cal Q}+1)+\left| {\cal Q}-1\right| ]}
 \left[(z+1)^{\left| {\cal Q}-1\right| }-1\right]}{H_0 \left| {\cal Q}-1\right|}.\\
 \label{GDE-power-r}
 \end{eqnarray}
%
% In order to plot the
To obtain the behavior of $\eta$ %versus
as a function of redshift parameter,
 %parameter, let us apply, for instance,
 we will apply the initial conditions %as
 $\eta(0)=0$
  and $d\eta(z)/dz\mid_{z=0}=0.1$, leading to
 % Therefore, equation \eqref{GDE-power-sol} gives
 \begin{eqnarray}\label{GDE-power-IC}
C_2=-C_1 =\frac{0.1}{\left|{\cal Q}-1\right| }.
\end{eqnarray}
Demanding $\alpha>1$, which corresponds
to an accelerating scale factor, from equation \eqref{AES-3}, we obtain
\begin{eqnarray}\label{acc-eq}
\frac{2}{(D-1){\cal W}\Gamma}>1.
\end{eqnarray}
Therefore, using \eqref{GDE-power-IC}, relations \eqref{GDE-power-sol} can be rewritten as
\begin{eqnarray}\label{eta-red}
\eta (z)=\frac{H_0}{10}r_0 (z)=\frac{(z+1)^{-1}-(z+1)^{-\frac{(D-1){\cal W}\Gamma}{2} }}{5
\left[(D-1){\cal W}\Gamma  -2\right]}.
%=\frac{\alpha \left[(z+1)^{\frac{\alpha-1}{\alpha }}-1\right]}{10 (\alpha -1) (z+1)}.
\end{eqnarray}

%In order to depict the behavior of $\eta (z)$ and $r_0 (z)$, let us
%obtain the allowed values of the parameters of the model
%by applying the energy conditions.
Setting $\rho=0=p$, and substituting $\rho_\phi$ and $p_\phi$ from
%relations
\eqref{Ex1-ro-phi} and \eqref{Ex1-p-phi} into
%inequalities
\eqref{NEC}-\eqref{DEC}, we obtain
%\begin{widetext}
\begin{eqnarray}\label{NEC2}
{\rm NEC:}\,\, \,\, &&{\cal W}\geq0,\\
\label{WEC2}
{\rm WEC:} \,\,\,\, &&\Gamma\geq0\hspace{33mm} {\rm and} \hspace{10mm} {\cal W}\geq0 ,\\
\label{SEC2}
{\rm SEC:} \,\,\,\, &&-\frac{2}{\Gamma}+(D-1){\cal W} \geq0 \hspace{12mm} {\rm and} \hspace{10mm} {\cal W}\geq0,\\
\label{DEC2}
{\rm DEC:}\,\, \,\,  &&\frac{2}{\Gamma}-{\cal W}\geq0,\hspace{8mm} \Gamma\geq0\hspace{8mm} {\rm and} \hspace{10mm} {\cal W}\geq0,
\end{eqnarray}
%\end{widetext}

where we have assumed $\phi^n>0$.

%Let us
%To determine the allowed values of $\Gamma$
%Respecting
Respecting the WEC and considering only an accelerating
scale factor at present, equations \eqref{acc-eq} and \eqref{WEC2} yield
%\rd{namely assuming}
%Concretely, assuming
%${\cal W}\geq0$ and $\Gamma\geq0$, %inequality
\begin{eqnarray}\label{acc-eq-2}
0\leq\Gamma {\cal W} <\frac{2}{(D-1)}.
\end{eqnarray}
%It is important to
Note that %choosing
for any value of ${\cal W}\Gamma$ that
satisfies \eqref{acc-eq-2}, the NEC and DEC are also satisfied, whilst the SEC is violated.

\begin{figure}
\centering\includegraphics[width=2.6in]{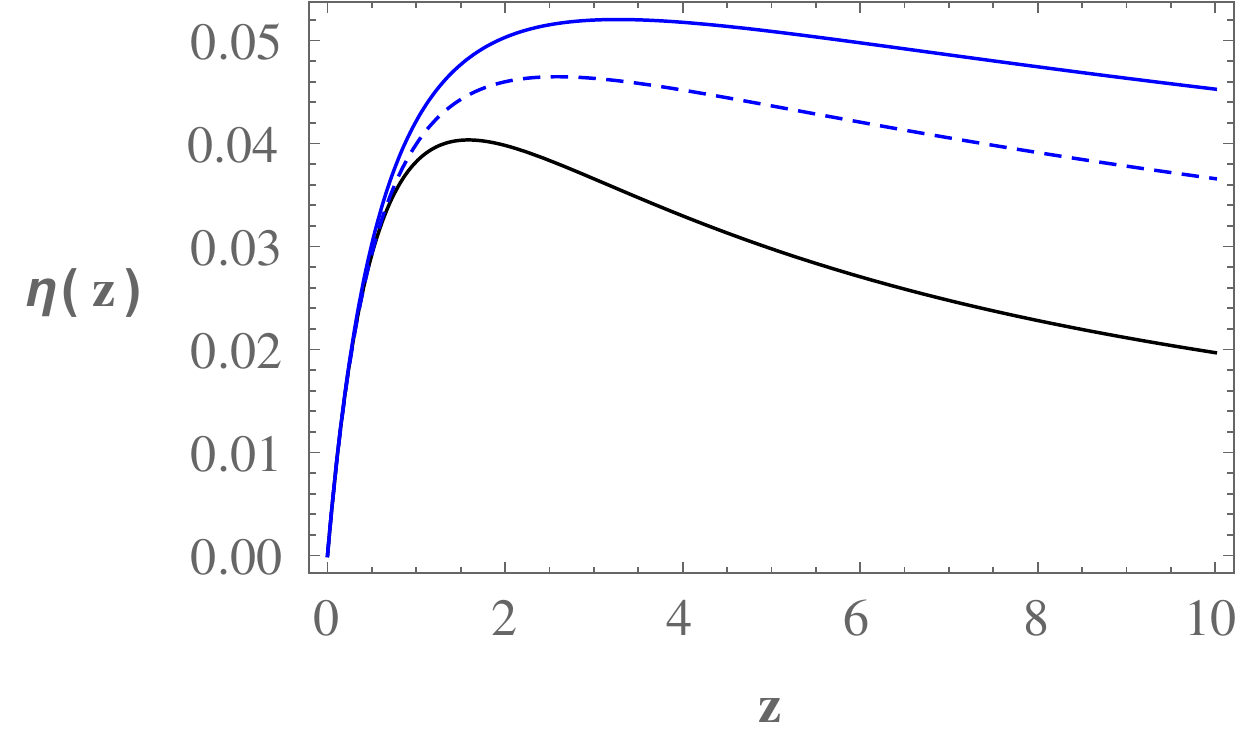}
\centering\includegraphics[width=2.6in]{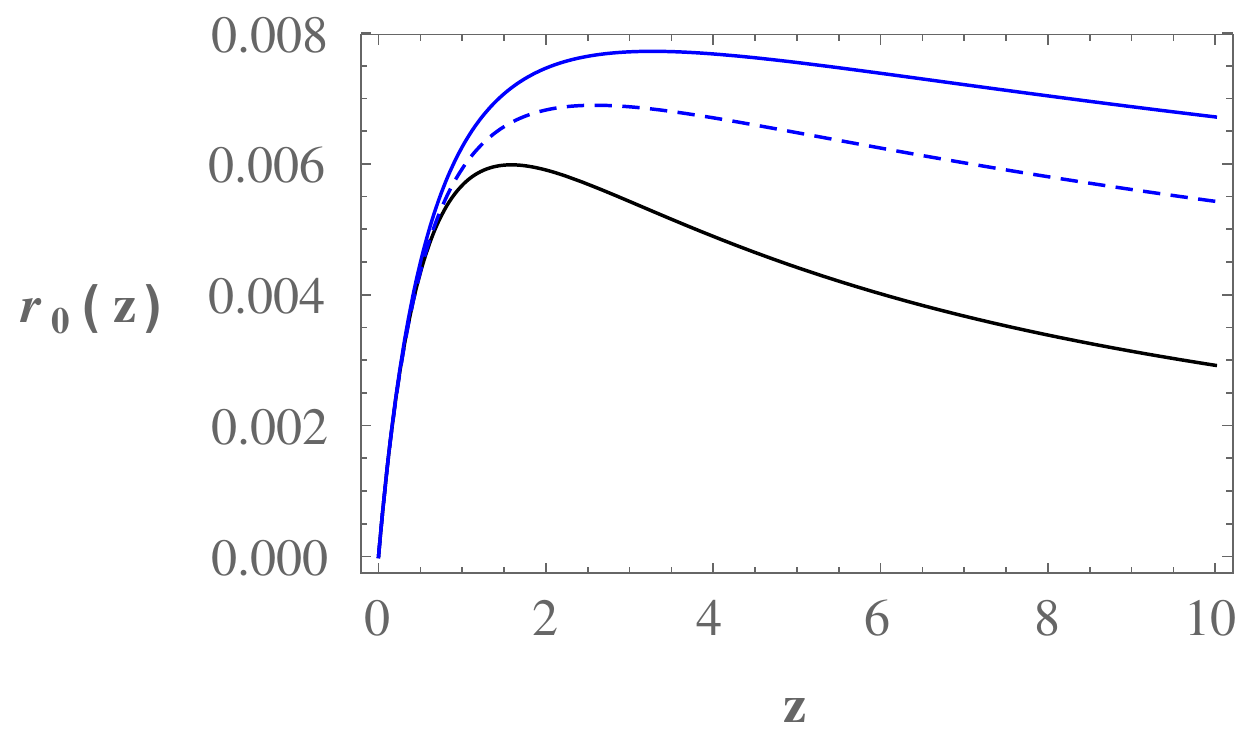}
\caption{{\footnotesize The behavior of $\eta(z)$ (the left panel)
and $r_0(z)$ (the right panel) associated with the null vector fields
with four dimensional FLRW background for the $\Lambda$CDM (the black curves)
and solution I (the solid and dashed blue curves).
We used the units of $H_0^{-1}$, $8 \pi G=1$, and we have
assumed $\eta(0)=0$, $d\eta(z)/dz\mid_{_{z=0}}=0.1$, $H_0=67.4 KM/s/Mps$.
For plotting the dashed and the solid curves, we have assumed ${\cal W}\Gamma=0.4$ ($q_0\simeq-0.4$)
and ${\cal W}\Gamma=0.3$ ($q_0\simeq-0.55$), respectively.
}}
%\foreignlanguage{english}
{\label{LM-eta-r}}
\end{figure}

Concretely, choosing the allowed values for ${\cal W}$ enables us to
obtain allowed values for $\Gamma$
(such that the inequality \eqref{acc-eq-2} is satisfied) in $D$ dimensions.
Therefore, we get the corresponding values for ${\cal Q}$ and
%finally we can completely depict
can depict the behavior of $\eta (z)$.
In figure \ref{LM-eta-r}, we show the
behavior of $\eta (z)$ and $r_0(z)$ for the allowed values of
the parameters associated with solution I and compare them with those of the $\Lambda$CDM model.

%%%%%%%%%%%%%%%%%%%%%%%%%%%%%%%%%%%%%%%%%%%%%%%%%%%%%%

\subsubsection{Solution II}
\label{SolII}
In this case, we would assume the potential $V(\phi)$ to be
\begin{eqnarray}\label{V-II-0}
V(\phi)=\frac{2}{\Gamma^2}\phi^n\dot{\phi}^4-{\cal W}\phi^n\dot{\phi}^2,
\end{eqnarray}
where $\Gamma>0$ is an arbitrary constant.
% with dimension $[\Gamma]=L^{-1}$.
Substituting $V(\phi)$ from \eqref{V-II-0} into \eqref{key-2}, we obtain
\begin{eqnarray}\label{V-II-1}
\left(\frac{1}{\phi^{\frac{3n}{2}}\dot{\phi}^4}\right)\frac{d}{dt}\left(\phi^n\dot{\phi}^4\right)=
-\kappa {\cal W} \Gamma\sqrt{2\left(\frac{D-1}{D-2}\right)}\equiv B,
\end{eqnarray}
where $|\phi^{\frac{n}{2}}\dot{\phi}^2|=\kappa\phi^{\frac{n}{2}}\dot{\phi}^2$ with $\kappa=\pm1$.

The above equation for arbitrary $n$ has an exact solution with a
complicated function. For simplicity of applying the GD equation
for this case, let us focus on the particular case where $n=0$ for which we get only one branch with $\kappa=+1$.
Therefore, it is easy to show that equation \eqref{V-II-1} yields
\begin{eqnarray}\label{phiDot-II}
\dot{\phi}(t)&=&\dot{\phi}_i {\rm Exp} \left[\frac{B}{4}(t-t_i)\right],\\
\label{phi-II}
\phi(t)&=&\frac{4\dot{\phi}_i}{B}\left\{{\rm Exp} \left[\frac{B}{4}(t-t_i)\right]-1\right\}
  +\phi_i,
\end{eqnarray}
where $t_i$, $\dot{\phi}_i$ and $\phi_i$ are integration
constants such that $\dot{\phi}(t_i)=\dot{\phi}_i$ and
${\phi}(t_i)=\phi_i$.
Moreover, substituting $V(\phi)$ from \eqref{V-II-0} into \eqref{key-2}, and
then applying \eqref{phiDot-II}, we obtain
\begin{eqnarray}\label{a-II}
a(t)=a_i \,{\rm Exp}\left\{\frac{2\dot{\phi}_i^2}{(D-1) {\cal W}\Gamma^2}
\left[1-
{\rm Exp} \left(\frac{B}{2}(t-t_i)\right)
\right]
\right\},
\end{eqnarray}
where $a_i$ is the value of the scale factor at arbitrary time $t=t_i$.

Furthermore, substituting the scalar field from \eqref{phi-II} into \eqref{V-II-0}, we get
\begin{eqnarray}\label{V-t-II}
V(t)=\frac{2\dot{\phi}_i^4}{\Gamma^2}{\rm Exp}\left[B(t-t_i)\right]
-\dot{\phi}_i^2{\cal W}{\rm Exp}\left[\frac{B}{2}(t-t_i)\right].
  \end{eqnarray}
Reemploying \eqref{phi-II}, we can obtain the scalar
potential in terms of $\phi$:
\begin{eqnarray}\label{V-phi-II}
V(\phi)=\frac{2\dot{\phi}_i^4}{\Gamma^2}
\left[1+\frac{B}{4\dot{\phi}_i}(\phi-\phi_i)\right]^4
-\dot{\phi}_i^2{\cal W}\left[1+\frac{B}{4\dot{\phi}_i}(\phi-\phi_i)\right]^2.
\end{eqnarray}
Substituting the scalar potential from \eqref{V-II-0} into
relations \eqref{rho-phi-gen} and \eqref{p-phi-gen}, we obtain
\begin{eqnarray}\label{rho-phi-II}
\rho_\phi(t)&=&\frac{\dot{\phi}_i^4}{\Gamma^2}{\rm Exp}\left[B(t-t_i)\right],\\
p_\phi(t)&=&{\cal W}\dot{\phi}_i^2{\rm Exp}\left[\frac{B}{2}(t-t_i)\right]
-\frac{\dot{\phi}_i^4}{\Gamma^2}{\rm Exp}\left[B(t-t_i)\right],
\label{p-phi-II}
\end{eqnarray}
where we used \eqref{phiDot-II}.
For this case, using relations \eqref{a-II}, \eqref{rho-phi-II}
and \eqref{p-phi-II}, it is straightforward to show that
the conservation law \eqref{cons-phi} is satisfied identically.
We should note that our herein formalism is the extended version of the
Barrow--Burd--Lancaster--Madsen model, see \cite{GC07} and references therein,
which for the particular case where ${\cal W}=1$ and $n=0$ reduces to their solution.

Let us now focus on the GD equation for the null vector fields associated with this case.
For the latter use, let us first obtain an important relation.
Assuming $t_i=t_0$,  $a_0=1$ and using equations \eqref{redshift} and \eqref{a-II}, we can easily show that
\begin{eqnarray}\label{t-z-rel}
{\rm Exp}\left[\frac{B}{2}(t-t_0)\right]=1+\left[\frac{(D-1){\cal W}\Gamma^2}{2\dot{\phi}_0^2}\right]{\rm ln}(1+z),
\end{eqnarray}
by which we can express all the quantities in terms of the redshift parameter.
Specifically, we can easily show that the Hubble parameter and the deceleration
parameter, $q\equiv-\ddot{a}/(a H^2)$, are
\begin{eqnarray}\label{H-II}
H(z)&=& \kappa {\cal W}\Gamma  \sqrt{\frac{D-1}{2 (D-2)}} \left[\frac{2 \dot{\phi}_0^2}{(D-1) {\cal W}\Gamma ^2 }+{\rm ln} (z+1)\right],\\\nonumber\\\nonumber\\
q(z)&=&-1+\frac{ \Gamma ^2 (D-1){\cal W} }{2\dot{\phi}_0^2}+\frac{(D-1)^2 {\cal W} ^2\Gamma ^4  {\rm ln} (z+1)}{4 \dot{\phi}_0^4}.
\label{q-II}
\end{eqnarray}
Assuming $\rho=0=p$, substituting the scale factor from \eqref{a-II} and the components of the EMT (of the SB
scalar field) from relations \eqref{rho-phi-II} and \eqref{p-phi-II} into equation \eqref{ops5}, and then using \eqref{t-z-rel}, we obtain
\begin{eqnarray}\label{GDE-II}
\frac{d^2\eta}{dz^2}+\left(\frac{{\cal P}}{1+z}\right)\frac{d\eta}{dz}
+\frac{{\cal Q}}{(1+z)^2}\eta=0,
\end{eqnarray}
 where
  \begin{eqnarray} \label{P-II}
  {\cal P}&=&\frac{8 \dot{\phi}_i^4+2(D-1) \dot{\phi}_i^2 {\cal W} \Gamma ^2
  +(D-1)^2 {\cal W} ^2\Gamma ^4 \, {\rm ln} (z+1)}{4 \dot{\phi}_i^4},
 \\
{\cal Q}&=&\frac{(D-1){\cal W}\Gamma ^2 }{2 \dot{\phi}_i^2+(D-1){\cal W}\Gamma ^2  \, {\rm ln} (z+1)}.
\label{Q-II}
\end{eqnarray}
We should note that among the quantities obtained
above, only $q$, ${\cal P}$ and ${\cal Q}$ do not depend on $\kappa$.
Nevertheless, equation \eqref{H-II} indicates that the
branch $\kappa=1$ is the physical one for solution II.

\begin{figure}
\centering\includegraphics[width=2.6in]{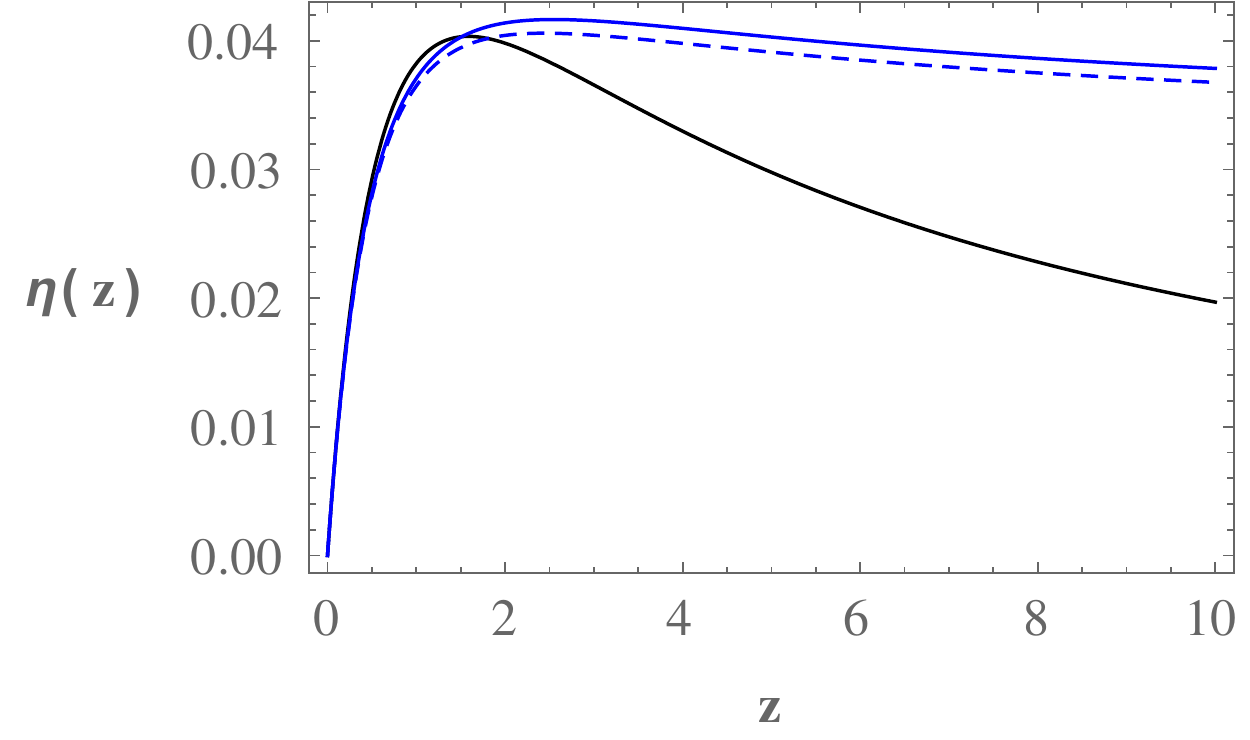}
\centering\includegraphics[width=2.6in]{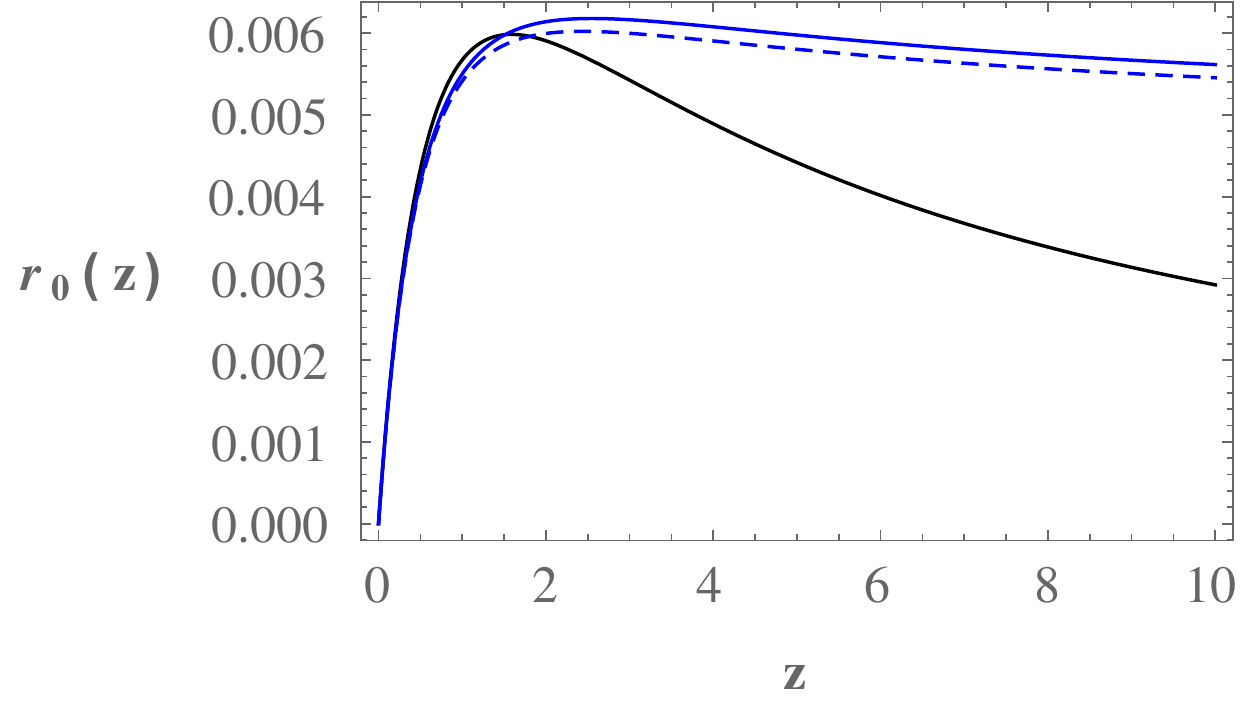}
\caption{{\footnotesize The behavior of $\eta(z)$ (the left panel)
and $r_0(z)$ (the right panel) associated with the null vector fields
with four dimensional FLRW background for the $\Lambda$CDM model (the black curves) and
the solution II (the dashed and solid blue curves). We used the units of $H_0^{-1}$, $8 \pi G=1$, and we have
assumed $\eta(0)=0$, $d\eta(z)/dz\mid_{_{z=0}}=1$, $H_0=67.4 km/s/Mps$.
For plotting the dashed curves (associated with a single scalar field minimally coupled to gravity, i.e., ${\cal W}=1$) and
the solid blue curves (associated with a SB model with ${\cal W}=0.95$), it has
been assumed $\mid\Gamma\mid=0.8$ and $\mid\dot{\phi}_i\mid=1.1$.
}}
%\foreignlanguage{english}
{\label{eta-r-II}}
\end{figure}

It is seen that for the
solution II, differently of the
solution I, ${\cal P}$ and ${\cal Q}$ are functions of
the redshift parameter $z$.
Therefore, it is realistically impossible to obtain an exact
solution for the differential equation \eqref{GDE-II} in a general case.
In this respect, let us analyze this solution by using numerical methods.
For this solution, assuming $\phi>0$, we see that the NEC and WEC
are satisfied provided that ${\cal W}\geq 0$.
Respecting the latter as well as admitting that our herein model might be suitable to
describe the accelerating universe at late times, we will
depict the behavior of $\eta(z)$ and $r_0(z)$ and compare them with those
 associated with the $\Lambda$CDM model, see, for instance, figure \ref{eta-r-II}.
 It is seen that for small values of $z$, the curves almost coincide.

 Substituting $V$ from \eqref{V-II-0}
into \eqref{SEC1}, we find that the SEC is satisfied provided that
   \begin{equation}\label{SEC-II-5}
\left \{
 \begin{array}{c}
  (D-1){\cal W}\Gamma ^2 \geq \frac{2 \dot{\phi}_i^2}{1-{\rm ln} (z+1)},    \hspace{10mm} {\rm for}\hspace{5mm} z < e-1 ,\\\\
(D-1){\cal W}\Gamma ^2 \leq \frac{2 \dot{\phi}_i^2}{1-{\rm ln} (z+1)},    \hspace{10mm} {\rm for}\hspace{5mm} z > e-1.
 \end{array}\right.
\end{equation}
On the other hand, demanding $q<0$, from \eqref{q-II}, we obtain
%\begin{widetext}
  \begin{eqnarray}\label{SEC-II-6}
   \frac{-1-\sqrt{1+ 4\dot{\phi}_i^2{\rm ln} (z+1)}}{{\rm ln} (z+1)}\leq(D-1){\cal W}\Gamma ^2 \leq \frac{-1+\sqrt{1+ 4\dot{\phi}_i^2{\rm ln} (z+1)}}{{\rm ln} (z+1)},  \hspace{7mm} {\rm for}\hspace{5mm} z>0.
\end{eqnarray}
%\end{widetext}
It is clear that the above determined regions for the
 quantity $(D-1){\cal W}\Gamma ^2$ given by inequalities
 \eqref{SEC-II-5} and \eqref{SEC-II-6} do not overlap for the corresponding values of $z$.
Therefore, by admitting $q<0$, the SEC is violated for the solution II.

%For instance, assuming $D=4$ and  we have depicted
%the behavior of the required quantities numerically.
%In.
%we have depicted the behavior of $\eta(z)$ and $r_0(z)$.

\subsection{Phantom dark energy model}
\label{Phantom}

Among various cosmological models, the simplest
dark energy model, i.e, $\Lambda$CDM model (the standard
cosmological scenario), definitely can provide
predictions, which are exquisitely well in agreement with the corresponding
observational data. Notwithstanding, with the enhancement of the number and
the accuracy of observations, it has been demonstrated that for some key cosmological
parameters estimated by $\Lambda$CDM model, there is still conspicuous
tension. Among them, the most obvious issue is estimating $H_0$ \cite{V21,V21-2, RCYBMZS21}.
More concretely, applying Planck cosmic microwave
background (CMB) and other cosmological observations
based on the $\Lambda$CDM model yields $H_0=67.364\pm 0.5km/s/Mpc$ \cite{PL.CosPar18}, which is much
smaller than that found by local measurements, particularly, with that estimated by
SH0ES collaboration by R20 team as $H_0=73.2\pm1.3km/s/Mpc$ at $\%68$ confidence
level \cite{RCYBMZS21} (in tension at $4.2\sigma$ with the Planck value
in a $\Lambda$CDM scenario \cite{V21}).

The above mentioned strong discrepancy in estimating the Hubble
constant has motivated scientific community to establish new physics beyond the concordance $\Lambda$CDM model to
reconcile or alleviate the $H_0$ tension. For instance, one of the most important
approaches has been introducing dynamical dark energy parameterizations scenarios for the
late times, for a detailed study of the well-known models, see \cite{V21}.
It has been demonstrated that most of these scenarios may solve the Hubble constant problem
 at the price of assuming a phantom-like dark energy equation of state \cite{V21,V21-2}.

A toy model of a {\it phantom} energy component, for which
$w_\phi<-1$ (where $w_\phi$ denotes the ratio of the pressure of the dark energy to its density), being
compatible with the observational data, has been established by Caldwell \cite{C02}.
A simple procedure to establish a phantom model is obtained by assuming the energy density and pressure
\eqref{rho-phi-gen} and \eqref{p-phi-gen} with negative kinetic term \cite{SS08}.
In this respect, in our herein model, we can also assume ${\cal W}<0$.
In addition to the phantom dark energy, concerning taking the form of
an ordinary matter associated with the present epoch, let us solve the field
equations \eqref{fun-Fri-1}-\eqref{cons-phi} in a particular
case where only the non-relativistic matter
 fills the universe, i.e., we assume $p=0$.
Therefore, equation \eqref{cons} yields
\begin{equation}\label{dust-ro}
\rho=\rho_0\left(\frac{a_0}{a}\right)^{D-1},
 \end{equation}
where the quantities with indices zero refer to their present values throughout.

As one of the main objectives of this paper is investigating the
GD equation for specific cosmological models, thus let us abstain from solving the generalized
complicated equations of motion associated with the herein phantom
model in the context of the extended SB framework.
Instead, we confine our attention to the most simplified phantom model
established by taking ${\cal W}=-1$, $n=0$ and $D=4$, which has been studied by
applying different procedures \cite{SS08,WZB10}.

In \cite{SS08}, by assuming a nearly
flat potential that satisfies slow-roll conditions, i.e.,
\begin{equation}\label{slow}
\left(\frac{V_{,\phi}}{V}\right)^2\ll1 \hspace{10mm} {\rm and} \hspace{10mm}\frac{V_{,\phi\phi}}{V}\ll1,
 \end{equation}
it has been shown
that the equation of state associated with the
scalar field, $w_\phi\equiv \frac{p_\phi}{\rho_\phi}$, is obtained
slightly less than $-1$ at present \cite{SS08}.
However, in \cite{WZB10}, by taking a reasonable assumption, analytic exact solutions
 have been obtained. In what follows, we focus on a phantom dark energy model
investigated in \cite{WZB10}, present more analysis of this model and finally investigate the GD equation for it.

It has been shown that for such a simple model (namely, assuming ${\cal W}=-1$, $n=0$ and $D=4$ ) the implicit
symmetries in the corresponding equations lead us to take an appropriate ansatz as \cite{WZB10}
\begin{equation}\label{phan-ansatz}
\dot{\phi}=-\sigma H,
 \end{equation}
where $\sigma>0$ is a constant. From equation \eqref{phan-ansatz}, we
obtain the evolution of the scale factor as
\begin{equation}\label{phan-a}
a=a_0 {\rm Exp}\left(\frac{\phi-\phi_0}{\sigma}\right),
 \end{equation}
 or equivalently, we get
\begin{equation}\label{phan-phi}
\phi=\phi_0 +\sigma\,{\rm ln}(1+z),
 \end{equation}
 where we set $a=1$ at the present.
Moreover, it is straightforward to show that \cite{WZB10}
\begin{eqnarray}\label{phan-H2}
H^2(\phi)\!\!&=&\!\!K_1\,{\rm Exp}\left(\frac{3\phi}{\sigma}\right)+K_2\,{\rm Exp}\left(-\sigma\phi\right),\\
\label{phan-V}
V(\phi)\!\!&=&\!\!-\sigma^2 K_1\,{\rm Exp}\left(\frac{3\phi}{\sigma}\right)+(6+\sigma^2)K_2\,{\rm Exp}\left(-\sigma\phi\right),
\\
\label{phan-rho-phi}
\rho_\phi\!\!&=&\!\!-K_1\sigma^2 \,{\rm Exp}\left(\frac{3\phi}{\sigma}\right)+3K_2\,{\rm Exp}\left(-\sigma\phi\right),\\\nonumber\\
\label{phan-p-phi}
p_\phi\!\!&=&\!\!-\left(3+\sigma^2\right)\,K_2\,{\rm Exp}(-\sigma\phi),
\end{eqnarray}
where $K_1$ and $K_2$ are integration constants.
Furthermore, from using equations \eqref{fun-Fri-1},
\eqref{phan-rho-phi} and \eqref{phan-p-phi}, we obtain the cosmological
density parameter associated with the ordinary matter as
\begin{eqnarray}\label{omega}
\Omega_{\rm m}=\frac{\rho}{3H^2}=\frac{K_1(3+\sigma^2)
{\rm Exp}\left(\frac{3\phi}{\sigma}\right)}{3K_1\,{\rm Exp}\left(\frac{3\phi}{\sigma}\right)
+3K_2\,{\rm Exp}\left(-\sigma\phi\right)},
\end{eqnarray}
which, according to equation \eqref{fun-Fri-1}, is related to the density parameter
associated with the phantom, $\Omega_{\phi}\equiv\frac{\rho_\phi}{3H^2}$, as
\begin{eqnarray}
\label{fri-eq}
\Omega_{\rm m}+\Omega_{\phi}=1.
\end{eqnarray}
It has been shown that the integration constants $K_1$ and $K_2$ are given by \cite{WZB10}
\begin{eqnarray}
\label{K1}
K_1=\frac{3H_0^2\left(1-\Omega_{\phi0}\right){\rm Exp}\left(-\frac{3\phi_0}{\sigma}\right)}{3+\sigma^2},\\\nonumber\\
\label{K2}
K_2=\frac{H_0^2\left(3\Omega_{\phi0}+\sigma^2\right){\rm Exp}\left(\sigma\phi_0\right)}{3+\sigma^2}.
\end{eqnarray}

For the later use for studying the GD equation \eqref{phan-GDE}, $H$ and $p_\phi$ should be
 expressed in terms of the redshift parameter z.
Substituting $\phi$, $K_1$ and $K_1$, respectively, from
\eqref{phan-phi}, \eqref{K1} and \eqref{K2} into \eqref{phan-H2}
and \eqref{phan-p-phi}, it is easy to show that
\begin{eqnarray}\label{Hz-phan}
H(z)&=&H_0\Bigg[\frac{3\Omega_{m0}}{3+\sigma^2}(1+z)^3+\left(\frac{3\Omega_{\phi0}+\sigma^2}
{3+\sigma^2}\right)(1+z)^{-\sigma^2}\Bigg]^{\frac{1}{2}},\\\nonumber\\
\label{p-phi-z}
p_{\phi}(z)&=&-H_0^2(3\Omega_{\phi0}+\sigma^2)(1+z)^{-\sigma^2}.
\end{eqnarray}
Moreover, using equation \eqref{ops3-1}, the deceleration parameter can be written as
\begin{eqnarray}
\label{q-phan-1}
q=-1+(1+z)\,\left(\frac{d\,{\ln}\left[ H(z)\right]}{dz}\right).
\end{eqnarray}
Then, substituting $H(z)$ from \eqref{Hz-phan} into \eqref{q-phan-1}, we obtain
\begin{eqnarray}\label{q-phan}
q(z)=-\frac{3+\sigma^2}{2}\left[\frac{3\Omega_{m0}(1+z)^{3+\sigma^2}}{3\Omega_{\phi0}+\sigma^2}+1\right]^{-1}+\frac{1}{2}.
\end{eqnarray}
It is seen that the amount of the deceleration parameter for a
 specific $z$ depends on the values taken by $\Omega_{\phi0}$ and
 the free parameter $\sigma$. Concretely, from \eqref{q-phan}, we see that the accelerating phase
 began only recently after a transition obtained
 from equation $q(z)=0$, which yields $z=z_{_{\rm Tr}}=f(\Omega_{\phi0}, \sigma)$.

Furthermore, using equations \eqref{phan-a}, \eqref{phan-phi} and \eqref{phan-rho-phi}--\eqref{K2}, it is
easy to show that $w_\phi$ can be written as
\begin{eqnarray}
\label{w-phan}
w_\phi=-\left(1+\frac{\sigma^2}{3\Omega_{\phi}}\right)=\frac{\left(\sigma^2+3\right)
\left(\sigma^2+3 \Omega_{\phi0}\right)}{3 \left(\sigma^2
+3 \Omega_{\phi0}\right)+3\sigma^2 (\Omega_{\phi0}-1) (z+1)^{\sigma^2+3}}.
\end{eqnarray}

In \cite{WZB10}, some important features of the
herein phantom dark energy model have been mentioned.
Nevertheless, in what follows, in addition to the GD equation, let
us further obtain a few novel interesting results.
We should note that, instead of the observational data used in \cite{WZB10},
%for the sake of investigating the Hubble constant problem explained at the begetting of the
 % present subsection,
let us focus on the considerations of \cite{PL.CosPar18,VMS16,CM20}, where we see
 $\{H_0=67.364\pm 0.5km/s/Mpc,\, \Omega_{m0}=0.315\pm0.007, w_{\phi0}=-1.03\pm 0.03\}$,
 $\{H_0=73.5.364\pm 2.5km/s/Mpc, w_{\phi0}=-1.29^{+ 0.15}_{-0.12}\}$
 and $\{H_0=75.35\pm 1.68km/s/Mpc,\, q_0=-1.08\pm0.29\}$, respectively.
 For instance, let us consider two examples: Assuming $\sigma=0.22$ ($1.06$) and $\Omega_{m0}=0.685$, the
 universe began acceleration very recently at redshifts about $z_{_{\rm Tr}}\simeq0.65$ ($0.76$).
 Moreover, we obtain $q_0\simeq-0.55$ ($-1.08$) and $w_{\phi0}\simeq-1.03$ ($-1.55$).
 It is seen that the results of the first example, disregarding the value of $w_{\phi0}$,
 is in agreement with the $\Lambda$CDM model \cite{PL.CosPar18}, while
 the second one (see the values in the parenthesis) is in
 agreement with the corresponding ones reported in \cite{CM20}.

\begin{figure}
\centering{}\includegraphics[width=2.6in]{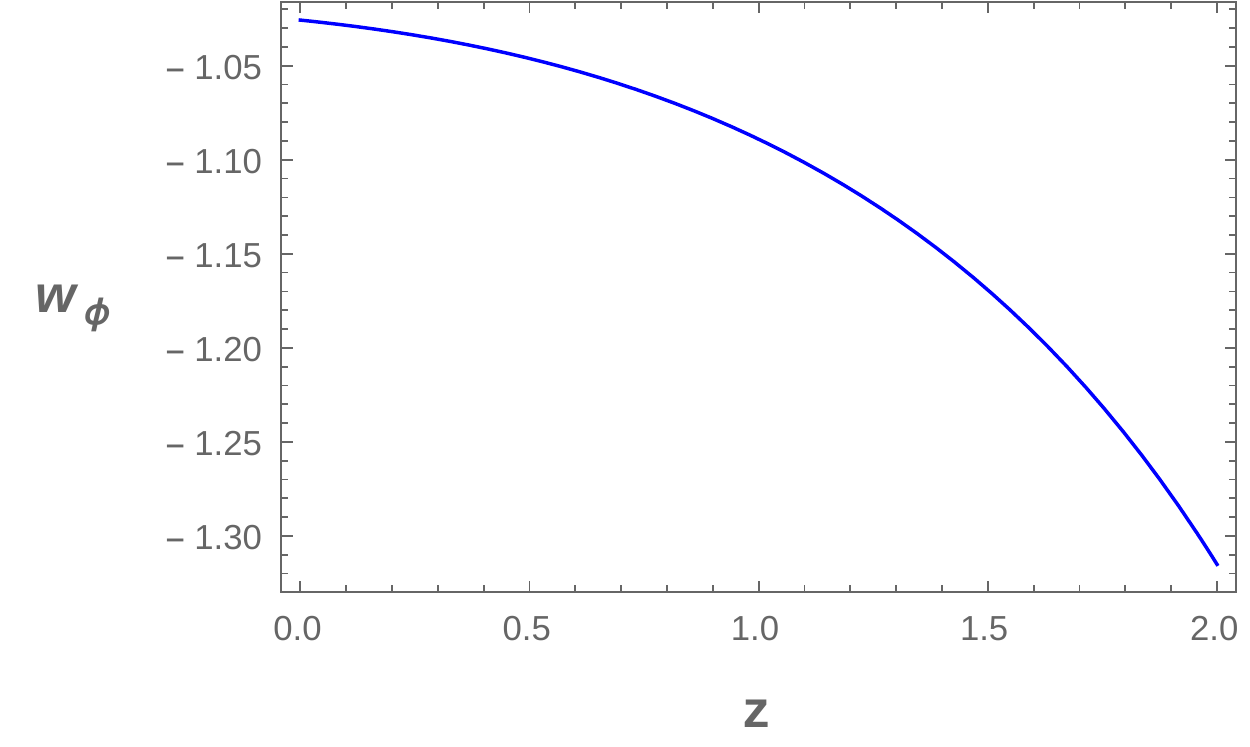}
\includegraphics[width=2.6in]{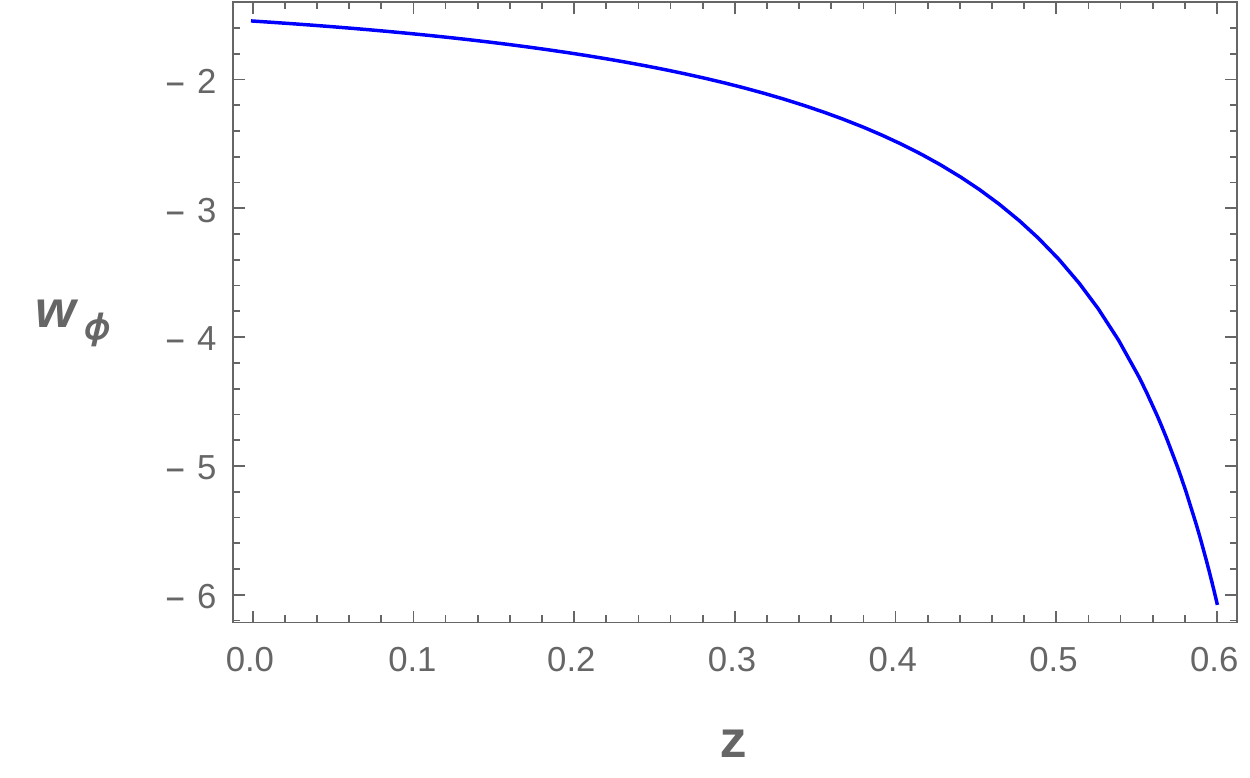}\hspace{3.2mm}
\caption{{\footnotesize The plot of $w_\phi(z)$ for small
values of the redshift parameter for our herein phantom model.
We considered a particular case of the SB theory with ${\cal W}=-1$ and $n=0$.
We have assumed $8 \pi G=1=c$, $\Omega _{\phi0}=0.685$, $\sigma=0.23$
(the left panel) and $\sigma=1.06$ (the right panel).
}}
%\foreignlanguage{english}
{\label{w-phantom}}
\end{figure}

First, let us plot the evolution of $w_\phi$ as function
of the redshift parameter, see figure \ref{w-phantom}.
We see that by choosing various values for the free
parameter $\sigma$, the model yields $w_{\phi0}$ such that it is in the
range reported by the recent observational data.

Secondly, before investigating the GD equation, we would study the late time
asymptotic behavior of some quantities as follows.

Substituting the integration constants from \eqref{K1} and \eqref{K2}
into \eqref{phan-V}, and then using \eqref{phan-phi}, we obtain
\begin{eqnarray}\label{V-phan-2}
V=\frac{H_0^2}{2 \left(\sigma ^2+3\right)} \Big[\left(6+\sigma ^2\right)
\left(3 \Omega_{ \phi 0}+\sigma ^2\right) (1+z)^{-\sigma ^2}
+3 \sigma ^2 (\Omega_{ \phi 0}-1) (1+z)^3\Big],
\end{eqnarray}
for which we get
\begin{eqnarray}
\label{V-phan-lim}
\lim_{z\to 0} \,V =\frac{1}{2} {H_0}^2 \left(\sigma ^2+6 \Omega_{\phi0}\right)={\rm constant}.
\end{eqnarray}
%Moreover, it is easy to show that
At late times, %the quantity
$V_{,\phi}/V$ also asymptotically approaches to a constant:
\begin{eqnarray} \label{V-phan-3}
\lim_{z\to 0} \,\frac{V_{,\phi}}{V} =\frac{\sigma  \left[9 (\Omega_{ \phi 0}-1)-\left(6+\sigma ^2\right)
 \left(3+\sigma ^2\Omega_{ \phi 0}\right)\right]}{\left(3+\sigma ^2\right) \left(6 \Omega_{ \phi 0}+\sigma ^2\right)}.
\end{eqnarray}
Furthermore, using relations \eqref{ops5-1}, we
obtain
\begin{eqnarray}
\label{w-phan-2}
\lim_{z\to 0} \,w_{\phi} =\frac{\sigma ^2}{3 (\Omega_{\phi0}-1)}-1\equiv w_{\phi0}={\rm constant},
\end{eqnarray}
which, as $\Omega_{\phi0}<1$, hence $w_{\phi0}$ will always be less than $-1$.

Let us now investigate the GD equation associated with this model.
Concerning our herein model, setting $p=0$ and $D=4$,
equation \eqref{ops5-1} reduces to
\begin{eqnarray}\label{phan-GDE}
\frac{d^2\eta}{dz^2}+\frac{1}{2(1+z)}
\left(7+\frac{ p_{\phi}}{H^2}\right)
\frac{d\eta}{dz}
+\frac{1}{2(1+z)^2}\left(3+\frac{p_\phi}{H^2}\right)\eta=0,
\end{eqnarray}

\begin{figure}
\centering\includegraphics[width=2.6in]{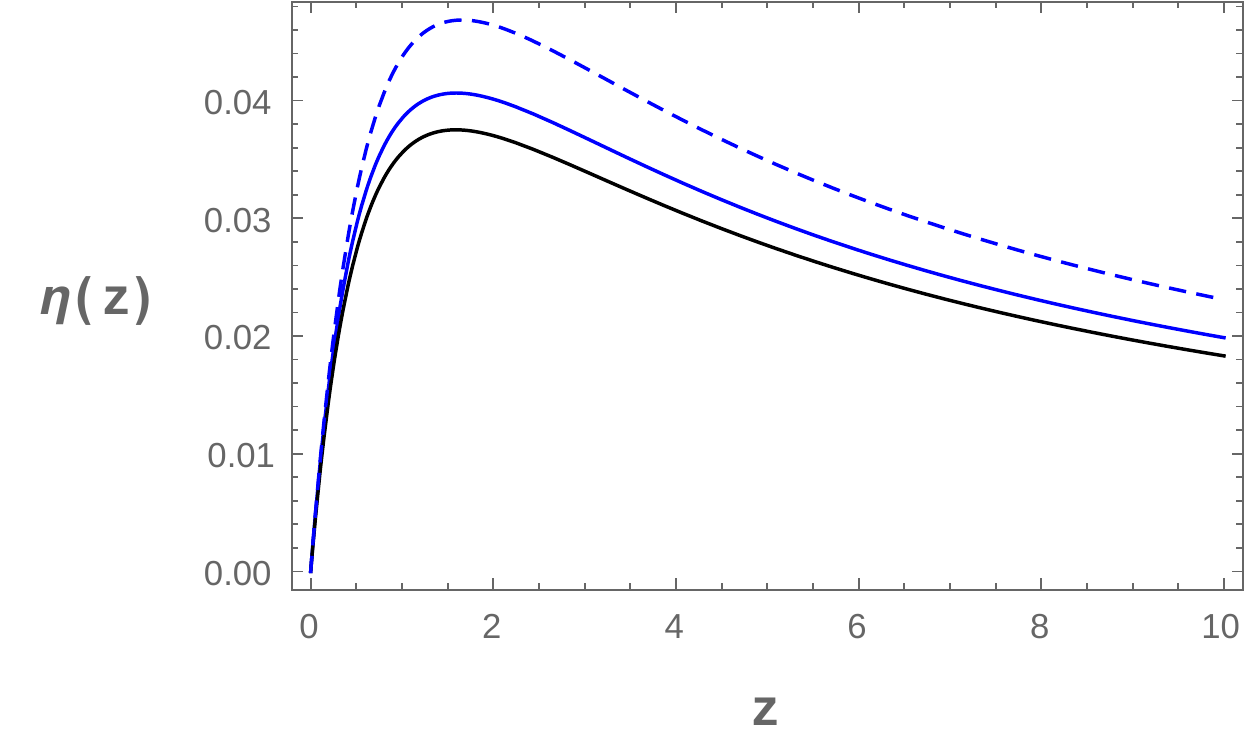}
\centering\includegraphics[width=2.6in]{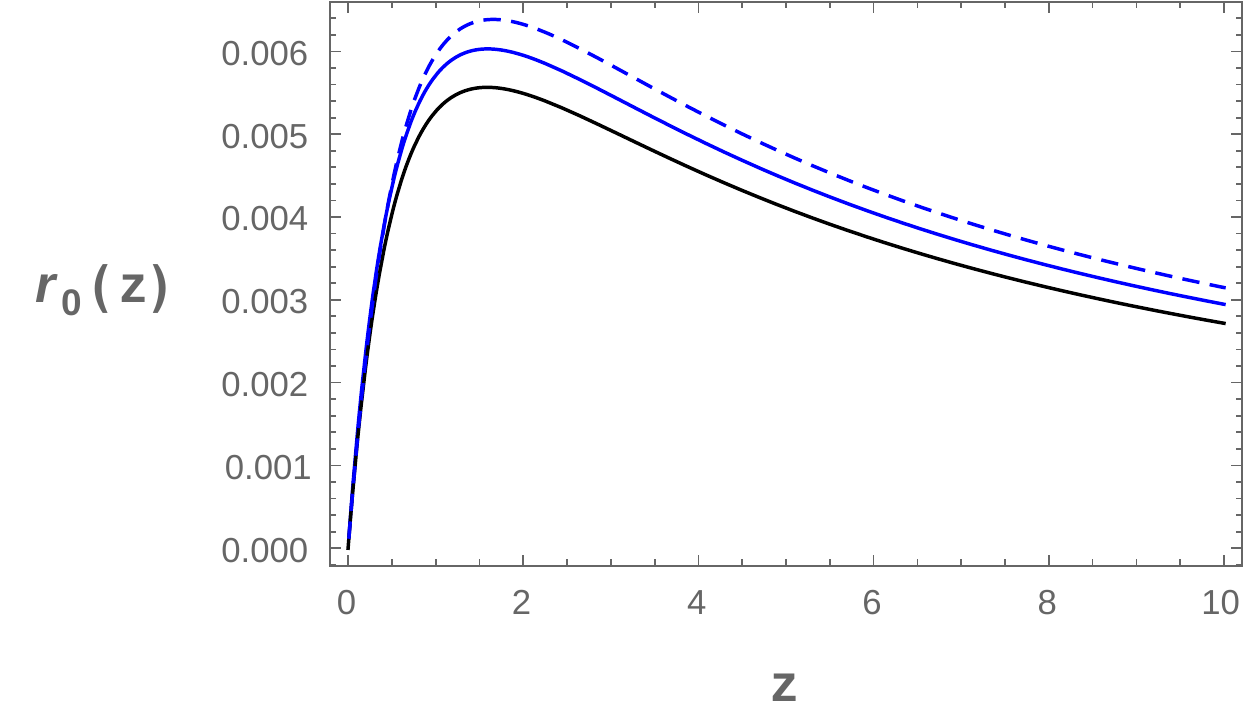}
\caption{{The behavior of $\eta(z)$ (the left panel)
and $r_0(z)$ (the right panel) associated with the null vector fields
with four dimensional FLRW background for the $\Lambda$CDM (the black curves) and phantom dark
energy model in the context of the SB theory for a
particular case where ${\cal W}=-1$, $n=0$ (the Blue curves).
We used units of $H_0^{-1}$, $8 \pi G=1=c$, and
assumed $\eta(0)=0$, $d\eta(z)/dz\mid_{_{z=0}}=0.1$,
$\Omega _{\Lambda }=0.685=\Omega_{\phi0}$ and $\Omega _{m0}=0.315$.
Furthermore, for plotting the black curves, solid blue
curves and dashed blue
curves, we used $H_0=67.4km/s/Mpc$, $\{H_0=67.4km/s/Mpc, q_0\simeq-0.55\, ({\rm or equivalently}\,\sigma=0.23)\}$
and $\{H_0=73.35km/s/Mpc, q_0\simeq-1.08 \,({\rm or equivalently} \,\sigma=1.06)\}$, respectively.
As the black curves and the solid blue curves almost coincide, for that we re-scale the former.
%\foreignlanguage{english}
{\label{eta-r-phantom}}
}}
%\foreignlanguage{english}
\end{figure}

Using equations \eqref{Hz-phan}-\eqref{q-phan}, one can show that
\begin{eqnarray}\label{H-q}
\left(\frac{H^2}{p_{\phi}}\right)^{-1}=2q(z)-1.
\end{eqnarray}
Therefore, the GD equation \eqref{phan-GDE} can be written as
\begin{eqnarray}\label{phan-GDE-2}
\frac{d^2\eta}{dz^2}+\left[\frac{3+q(z)}{(1+z)}\right]\frac{d\eta}{dz}
+\left[\frac{1+q(z)}{(1+z)^2}\right]\eta=0,
\end{eqnarray}
where $q(z)$ is given by \eqref{q-phan}.
It seems that it is not possible to obtain analytic exact solutions for \eqref{phan-GDE-2}.
In this regard, we will use a numerical approach to analyze it.
Using recent observational data \cite{PL.CosPar18}, in figure \ref{eta-r-phantom},
we plot $\eta$ and $r_0$ against the redshift parameter (see the blue solid curves).
In addition to the observational data reported in \cite{PL.CosPar18}, the blue dashed curves
show the behavior of $\eta(z)$ and $r_0(z)$ by
considering the value of the $H_0$ estimated by SH0ES collaboration, see
for instance, \cite{RCYBMZS21}.
In these figures, we have also compared the behavior of $\eta$ and $r_0$ associated with herein
phantom dark energy model with the corresponding case (i.e., assuming $p=0$)
in the $\Lambda$CDM model presented in subsection \ref{GR}.
It is seen that the general behavior of $\eta$ and $r_0$ are similar for all, as expected.

\section{GD equation in the context of the MSBT}
\label{MSBT}

In this section, let us first review the MSBT in arbitrary dimensions~\cite{RM19,RPSM20}, and then
investigate the GD equation in this framework.

%The action associated with a $(D+1)$-dimensional SB theory in the
%absence of the scalar potential, in analogy with the
%corresponding four-dimensional case~\cite{SB85-original}, is given by
%\begin{equation}\label{SB-5action}
%{\cal S}^{^{(D+1)}}=\int d^{D+1}x \sqrt{\Bigl|{}{\cal
%G}^{^{(D+1)}}\Bigr|} \,\left[R^{^{(D+1)}}-{\cal W}\phi^n\,{\cal
%G}^{ab}\,(\overline{\nabla}_a\phi)(\overline{\nabla}_b\phi)+\chi\,
%L\!^{^{(D+1)}}_{_{\rm matt}}\right],
%\end{equation}
%where ${\cal G}^{^{(D+1)}}$ and $R^{^{(D+1)}}$ stand for the determinant
%and Ricci scalar associated with the
%$(D+1)$-dimensional metric ${\cal G}_{{ab}}$, respectively; $\overline{\nabla}$ denotes the covariant derivative on the $(D+1)$-dimensional space-time %(bulk).
 %$\chi=8\pi$, $\phi$ is a scalar field (hereafter we call it the SB scalar field), and ${\cal W}$, $n$ are two dimensionless parameters of the % %model\rlap.\footnote{\bl{In our herein work, we have been using the units where $G=1=c$ (where $G$ and $c$ are the Newton gravitational constant and the %speed of light, respectively). The Latin and Greek indices run from zero to $D$.}} In contrast to the original non-compact Kaluza-Klein setting of %IMT~\cite{PW92,OW97}, in \cite{RPSM20}, in order to establish a more
%generalized setting, the Lagrangian associated with the ordinary matter fields (which is independent of the SB scalar field) has also been assumed in the %bulk, i.e.,  $L^{^{(D+1)}}_{_{\rm matt}}\neq0$.

In analogy to \eqref{BD-Eq-DD} and \eqref{D2-phi}, their $(D+1)$-dimensional counterpart field
equations, in the absence of the scalar potential, are given by
\begin{eqnarray}\label{(D+1)-equation-1}
G^{^{(D+1)}}_{ab}={\cal W}\phi^{n}\left[(\overline{\nabla}_a\phi)(\overline{\nabla}_b\phi)
-\frac{1}{2}{\cal G}_{ab}(\overline{\nabla}^c\phi)(\overline{\nabla}_c\phi)\right]
+T^{^{(D+1)}}_{ab}
\end{eqnarray}
and
\begin{equation}\label{(D+1)-equation-2}
2\phi^n\overline{\nabla}^2\phi
+n\phi^{n-1}(\overline{\nabla}_a\phi)(\overline{\nabla}^a\phi)=0,
\end{equation}
where $\overline{\nabla}$ is the covariant derivative
associated with $(D+1)$-dimensional spacetime (bulk) and
$\overline{\nabla}^2\equiv\overline{\nabla}_a\overline{\nabla}^a$.
We should note that the Lagrangian associated with the ordinary matter fields has also
been taken nonzero in the bulk, i.e.,  $L^{^{(D+1)}}_{_{\rm matt}}\neq0$.
This choice was made in \cite{RPSM20} with the purpose to establish a more generalized setting.
Moreover, the tensors and quantities with index $(D+1)$ and/or Latin
indices (the Latin indices run from zero to $D$)
are also associated with the $(D+1)$-dimensional spacetime (bulk).

%%%%%%%%%%%%%%%%%%%%%%%%%%%%%%%%%%%%%%%%%%%%%%%%%%%%%%%%%%%%%%%%%%%%%%%%%%%%%%%%%%%%
%%%%%%%%%%%%%%%%%%%%%%%%%%%%%%%%%%%%%%%%%%%%%%%%%%%%%%%%%%%%%%%%%%%%%%%%%%%%%%%%%%%%%
%%%%%%%%%%%%%%%%%%%%%%%%%%%%%%%%%%%%%%%%%%%%%%%%%%%%%%%%%%%%%%%%%%%%%%%%%%%%%%%%%%%%%%
Applying a specific reduction procedure, and considering \cite{PW92,OW97},
\begin{eqnarray}\label{global-metric}
dS^{2}={\cal G}_{ab}(x^c)dx^{a}dx^{b}=g_{\mu\nu}(x^\alpha,l)dx^{\mu}dx^{\nu}+
\epsilon\psi^2\left(x^\alpha,l\right)dl^{2},
\end{eqnarray}
it has then been shown that the effective EMT as well as an induced scalar
potential emerge intrinsically from the geometry of the extra
dimension (for more detail, see \cite{RM19,RPSM20}).
In \eqref{global-metric}, $l$ denotes
a non-compact coordinate along the extra dimension;
the scalar field $\psi$ depends on all coordinates and $\epsilon=\pm1$.
The hypersurface $\Sigma_0$ corresponding to $l=l_{0}={\rm constant}$ is orthogonal to the
$(D+1)$-dimensional unit vector
\begin{equation}\label{unitvector}
n^a=\frac{\delta^a_{_D}}{\psi}, \qquad {\rm where} \qquad
n_an^a=\epsilon,
\end{equation}
along the extra dimension.
Therefore, the induced metric $g_{\mu\nu}$ on the hypersurface
$\Sigma_{0}$ is given by
\begin{equation}\label{brane-metric}
ds^{2}={\cal G}_{\mu\nu}(x^{\alpha},
l_{0})dx^{\mu}dx^{\nu}\equiv g_{\mu\nu}dx^{\mu}dx^{\nu}.
\end{equation}
Consequently, four sets of equations are retrieved (see \cite{RPSM20} where more details can be found):
\begin{enumerate}
\item
An equation for the scalar field $\psi$:
\begin{eqnarray}\label{D2say}
\frac{{\nabla}^2\psi}{\psi}=-\frac{\epsilon}{2\psi^2}
\left[g^{\lambda\beta}\overset{**}{g}_{\lambda\beta}
+\frac{1}{2}\overset{*}{g}^{\lambda\beta}\overset{*}{g}_{\lambda\beta}
-\frac{g^{\lambda\beta}\overset{*}{g}_{\lambda\beta}\overset{*}{\psi}}{\psi}\right]
-\frac{\epsilon{\cal W}\phi^n(\overset{*}{\phi})^2}{\psi^2}
+\left[\frac{T^{^{(D+1)}}}{D-1}-\frac{\epsilon T^{^{(D+1)}}_{_{DD}}}{\psi^2}\right],
\end{eqnarray}
where $\overset{*}A\equiv\frac{\partial A}{\partial l}$.

\item
The counterpart of the conservation law presented in IMT is given by
\begin{eqnarray}\label{G-D,alpha}
G_{\alpha D}^{^{(D+1)}}&=&R_{\alpha D}^{^{(D+1)}}=\psi P^{\beta}{}_{\alpha;\beta}
= T^{^{(5)}}_{\alpha 4}+{\cal W}\phi^n\overset{*}{\phi}({\nabla}_\alpha\phi),
\end{eqnarray}
where
\begin{equation}\label{P-mono}
P_{\alpha\beta}\equiv\frac{1}{2
\psi}\left(\overset{*}{g}_{\alpha\beta}
-g_{\alpha\beta}g^{\mu\nu}\overset{*}{g}_{\mu\nu}\right).
\end{equation}

\item
The second pair of the field equations associated
with the MSBT are also given by \eqref{BD-Eq-DD} and \eqref{D2-phi}.
However, in contrary to the conventional SB theory presented in section \ref{SetUp}, in the MSBT
framework the EMT as well as scalar potential are {\it not} added by phenomenological assumptions, but
instead they are fully emerge from the geometry. Concretely,
(i) the induced scalar potential $V(\phi)$ is obtained from
\begin{eqnarray}\label{v-def}
 V_{,\phi}\equiv-\frac{2{\cal W}\phi^n}{\psi^2}
\Bigg\{\psi({\nabla}_\alpha \psi)({\nabla}^\alpha\phi)
+\frac{n\epsilon}{2}\left(\frac{\overset{*}{\phi}^2}{\phi}\right)
+\epsilon\left[\overset{**}{\phi}+\overset{*}{\phi}
\left(\frac{g^{\mu\nu}\overset{*}{g}_{\mu\nu}}{2}
-\frac{\overset{*}{\psi}}{\psi}\right)\right]\Bigg\}.
\end{eqnarray}
(ii) The induced EMT, $T_{\mu\nu}^{^{(D)}}$,
in \eqref{BD-Eq-DD} has four terms:
\begin{eqnarray}\label{IM}
 T_{\mu\nu}^{^{(D)}}= E_{\mu\nu}+T_{\mu\nu}^{^{[\rm IMT]}}+T_{\mu\nu}^{^{[\rm \phi]}},
+\frac{1}{2}g_{\mu\nu}V(\phi)
\end{eqnarray}
where
\begin{itemize}
\item
$E_{\mu\nu}$ represents the effective EMT induced from the
$(D+1)$-dimensional ordinary matter fields assumed in the bulk:
\begin{eqnarray}\label{S}
E_{\mu\nu}\equiv T_{\mu\nu}^{^{(D+1)}}-
g_{\mu\nu}\left[\frac{T^{^{(D+1)}}}{D-1}-
\frac{\epsilon\, T_{_{DD}}^{^{(D+1)}}}{\psi^2}\right].
\end{eqnarray}

\item

$T_{\mu\nu}^{^{[\rm IMT]}}$ is the same induced matter presented in the IMT \cite{PW92}:
\begin{eqnarray}\nonumber
T_{\mu\nu}^{^{[\rm IMT]}}&\!\!\!\equiv &\!\!
\frac{{\nabla}_\mu{\nabla}_\nu\psi}{\psi}
-\frac{\epsilon}{2\psi^{2}}\Bigg\{\frac{\overset{*}{\psi}\overset{*}{g}_{\mu\nu}}{\psi}-\overset{**}{g}_{\mu\nu}
+g^{\lambda\alpha}\overset{*}{g}_{\mu\lambda}\overset{*}{g}_{\nu\alpha}-\frac{1}{2}
g^{\alpha\beta}\overset{*}{g}_{\alpha\beta}\overset{*}{g}_{\mu\nu}\\
 \!\!\!&+&\!\!\frac{\epsilon g_{\mu\nu}}{4}
\left[\overset{*}{g}^{\alpha\beta}\overset{*}{g}_{\alpha\beta}
+\left(g^{\alpha\beta}\overset{*}{g}_{\alpha\beta}\right)^{2}\right]\Bigg\}.
\label{IMTmatt.def}
\end{eqnarray}

\item

Another term of the induced EMT is:
\begin{eqnarray}
\label{T-phi} T_{\mu\nu}^{^{[\rm
\phi]}}\equiv
\left[\frac{\epsilon{\cal W}}{2}\phi^n\left(\frac{\overset{*}{\phi}}{\psi}\right)^2\right]g_{\mu\nu}.
\end{eqnarray}

\end{itemize}

\end{enumerate}

In summary, by considering the metric (\ref{global-metric})
and selecting a dimensional
reduction procedure, the equations (\ref{(D+1)-equation-1}) and (\ref{(D+1)-equation-2})
associated with the $(D+1)$-dimensional SB theory
(in the absence of any potential and cosmological constant), are then
reduced to the effective field equations~(\ref{BD-Eq-DD}), (\ref{D2-phi}), (\ref{D2say})
and (\ref{G-D,alpha}) on the hypersurface.
From the viewpoint of an observer on the hypersurface (who has no information concerning the
 reduction procedure as well as the existence of the extra dimension),
 ~(\ref{BD-Eq-DD}) and~(\ref{D2-phi}) would be considered as the field equations for the
SB theory (with a scalar potential) in $D$ dimensions, which
can also be derived from the action \eqref{induced-action} admitting
\begin{equation}\label{induced-source}
\sqrt{-g}\left(E_{\alpha\beta}+T^{^{[\rm
SB]}}_{\alpha\beta}\right)\equiv 2\delta\left( \sqrt{-g}\,
L\!^{^{(D)}}_{_{\rm matt}}\right)/\delta g^{\alpha\beta}.
 \end{equation}

In order to proceed our considerations in the context of the MSBT, let us mention an important remark:
%important point:
Equations (\ref{BD-Eq-DD}) and (\ref{D2-phi}) are the field
equations that are valid not only for the conventional SB theory but also for the
MSBT. However, concerning the former, both the EMT and the scalar potential
should be chosen from phenomenological assumptions.
Whilst for the latter, not only the EMT but also the scalar
potential are thus extracted from the corresponding equations,
namely, equations \eqref{IM} and \eqref{v-def}. More concretely,
for the MSBT, we will employ the EMT as well as scalar
potential directly dictated from the geometry.

\subsection{GD equation for null vector field for cosmological models in the MSBT theory}

In what follows, let us focus our attention on the GD equation in context of the MSBT.
We should note that in order to obtain
equation \eqref{Rie-ten-2}, we have merely used
equations (\ref{BD-Eq-DD}) and (\ref{D2-phi}) without imposing any
constraint.  Therefore, it is also valid when we take MSBT as the underlying theory.

Nevertheless, %regarding the subsequent stage,
it is worthy to
stress that equation \eqref{GDE-gen-0} has been deduced for the
special case where we restricted ourselves to a
perfect fluid. Therefore, this equation will be valid within the MSBT provided that
 the geometrically induced matter (on a $D$-dimensional
hypersurface) to be also a perfect fluid. In this
respect, let us choose the same
assumptions used to derive the GD equation.
More concretely, we consider a
particular case of the metric \eqref{global-metric}:
\begin{equation}\label{bulk-FRW}
dS^{2}=ds^2+\epsilon \psi^2(t)dl^{2},
\end{equation}
where the line-element associated with the hypersurface is
given by \eqref{ohanlon metric-2}. Moreover, we assume that
there is no ordinary matter fields in the bulk.
Therefore, from equations \eqref{S}, we get $E_{\mu\nu}=0$.
%On the other hand,
By imposing the cylinder
condition \cite{OW97} (by which we must set the derivatives with respect to $l$
equal to zero), from equation \eqref{T-phi}, we obtain $T_{\mu\nu}^{^{[\rm\phi]}}=0$.
%Furthermore, admitting above mentioned
Regarding the assumptions mentioned above,
equations \eqref{IM}, \eqref{D2say}, \eqref{v-def}
and \eqref{IMTmatt.def} reduce to
\begin{eqnarray}\label{cylinder-1}
 T_{\mu\nu}\!\!&=&\!\!\frac{\nabla_\mu\nabla_\nu\psi}{\psi}
 +\frac{V(\phi)}{2}g_{\mu\nu}, \hspace{5mm}
T=\frac{D  V(\phi)}{2 },\\\nonumber\\\nonumber
\\
V_{,\phi}\!\!&\equiv &\!\! -\frac{2{\cal W}\phi^n}{\psi}
({\nabla}_\alpha \psi)({\nabla}^\alpha\phi)
, \hspace{5mm} \nabla^2\psi=0.
\label{cylinder-2}
\end{eqnarray}
%Consequently, by
Assuming $\phi=\phi(t)$ and substituting the
components of the metric \eqref{ohanlon metric-2} into \eqref{IM},
the energy density $\rho$ and pressure $p$ of the induced matter is given by
\begin{eqnarray}\label{R16}
\rho \!\!\!& \equiv \!\!\!& - T^{0}_{\,\,\,0}=\frac{\ddot{\psi}}{\psi}-\frac{V(\phi)}{2},\\\nonumber
\\
\label{R17}
p \!\!\!& \equiv \!\!\!&  T^{i}_{\,\,\,i}=
-\frac{\dot{a}\dot{\psi}}{a\psi}+\frac{V(\phi)}{2},
\end{eqnarray}
where $i=1,2, ..., (D-1)$ (with no sum on $i$). Moreover, $V(\phi)$ in
relations \eqref{R16} and \eqref{R17} should be obtained from solving the
 differential equation (\ref{cylinder-2}):
\begin{equation}\label{new30}
V_{,\phi}{\Big|}_{_{\Sigma_{o}}}\!\!\!\!\!=
2{\cal W}\phi^n\dot{\phi}\left(\frac{\dot{\psi}}{\psi}\right).
\end{equation}

Hence, \eqref{R17} implies
that the pressure in all directions are equal (i.e., $p_1=p_2=p_i\equiv p$), and consequently the
induced matter on the $D$-dimensional hypersurface is a perfect fluid.
The induced matter also obeys \eqref{perfect}. So, we conclude
that \eqref{GDE-gen-0}, \eqref{FT-SB} and \eqref{GDE-gen-2}
(which have been deduced in the SB framework)
can also be applicable within the MSBT.
However, let us emphasize once again that, contrary to the standard SB theory, herein $\rho$, $p$
and $V(\phi)$ have not put %substituted
by hand, but instead
they emerge from the geometry of the higher dimensions.

%In order to proceed towards the

To study the GD equation in the context of the MSBT, let us first obtain
exact solutions of our herein cosmological model (for more detail, see \cite{RPSM20}).
Equations \eqref{(D+1)-equation-2} and  \eqref{cylinder-2},
respectively, lead us to the following constants of motion
\begin{eqnarray}\label{new1}
a^{D-1}\phi^{\frac{n}{2}}\dot{\phi}\psi=c_1,\\\nonumber\\
\label{34}
a^{D-1}\dot{\psi}=c_2,
\end{eqnarray}
where $c_1\neq0$ and $c_2\neq0$ are constants of integration.
Equations \eqref{new1} and \eqref{34} imply
%lead us to the following relations
\begin{equation}\label{new7}
\psi=\left \{
 \begin{array}{c}
  \psi_i {\rm Exp}\left(\frac{2\beta}{n+2}\phi^{\frac{n+2}{2}}\right)
 \hspace{21mm} {\rm for}\hspace{5mm} n\neq-2,\\\\
 \psi_i\phi^\beta
  \hspace{37mm} {\rm for}\hspace{5mm} n=-2,
 \end{array}\right.
\end{equation}
and
\begin{equation}\label{new17}
a=\left \{
 \begin{array}{c}
  a_i {\rm Exp}\left[\frac{2\gamma(D)}{n+2}\phi^{\frac{n+2}{2}}\right]
 \hspace{19mm} {\rm for}\hspace{5mm} n\neq-2,\\\\
 a_i\phi^{\gamma(D)}
  \hspace{35mm} {\rm for}\hspace{5mm} n=-2,
 \end{array}\right.
\end{equation}
where to obtain \eqref{new17}, we have also used the Friedmann equation
associated with the bulk in the absence of the ordinary matter \cite{RPSM20}.
Moreover, $\psi_i$ and $a_i$ are constants of
integration, $\beta\equiv\frac{c_2}{c_1}$ and $\gamma(D)$ was defined as
\begin{eqnarray}\label{new17-2}
\gamma(D)\equiv\frac{1}{D-2}\left[-\beta\pm\sqrt{\beta^2+\left(\frac{D-2}{D-1}\right){\cal W}}\right].
\end{eqnarray}
Replacing $\psi$ and $a$ from relations \eqref{new7} and \eqref{new17} into equation \eqref{new1}, we get
\begin{equation}\label{new18-2}
\left \{
 \begin{array}{c}
 \dot{\phi}\phi^{\frac{n}{2}}{\rm Exp}\left[\frac{2f(D)}{n+2}
 \phi^{\frac{n+2}{2}}\right]=\frac{c_1a_i^{1-D}}{\psi_i}
 \hspace{15mm} {\rm for}\hspace{5mm} n\neq-2,\\\\\\
   \dot{\phi}\phi^{f(D)}=\frac{c_1a_i^{1-D}}{\psi_i}
  \hspace{35mm} {\rm for}\hspace{5mm} n=-2,
 \end{array}\right.
\end{equation}
where
\begin{eqnarray}\label{f}
f(D)\equiv (D-1)\gamma(D)+\beta.
\end{eqnarray}
In order to obtain the unknowns of the model in terms of the cosmic
time, we should first obtain $\phi(t)$ by solving the above differential equations.
However, whether or not $f(D)$ is chosen to vanish, we obtain two classes of exact solutions.
%In what follows, let us merely summarize the results obtained
%in \cite{RPSM20} and abstain from repeating the procedure.

\subsubsection{GD equation for Exponential-law solution}

In the particular case where $f(D)=0$, the exact solutions
corresponding to the equations~(\ref{new18-2}) are given by \cite{RPSM20}
\begin{equation}  \label{new54}
 \phi(t)=\left \{
 \begin{array}{c}
\left[\frac{(n+2)(1-D)h(D)(t-t_i)}{2 \beta}\right]^{\frac{2}{n+2}}
 \hspace{8mm} {\rm for}\hspace{5mm} n\neq-2,\\\\
  {\rm Exp}\left[\frac{(1-D)h(D)(t-t_i)}{\beta}\right]
  \hspace{14mm} {\rm for}\hspace{5mm} n=-2,
  \end{array}\right.
\end{equation}
where $t_i$ is an integration constant and
\begin{equation}\label{H-d}
h(D)\equiv \frac{c_1\beta a_i^{1-D}}{(1-D)\psi_i}.
\end{equation}
%Let us summarize the other relations obtained in
From \cite{RPSM20}:
\begin{eqnarray}\label{new59}
 a(t)\!\!&=&\!\!a_i\,{\rm Exp}\left[h(D)\left(t-t_i\right)\right],\hspace{14mm} \forall n,\\\nonumber\\
 \label{new59-2}
 \psi(t)\!\!&=&\!\!\psi_i\,{\rm Exp}\left[(1-D) h(D)(t-t_i)\right],\hspace{4mm} \forall n.
\end{eqnarray}
%Moreover,
The induced potential on the hypersurface is:
\begin{equation}\label{new66}
V(\phi)=\left \{
 \begin{array}{c}
\frac{2V_0}{n+2}\,\phi^{\frac{n+2}{2}}
 \hspace{17mm} {\rm for}\hspace{8mm} n\neq-2,\\\\
 V_0\,{\rm ln}\left(\frac{\phi}{\phi_i}\right)
 \hspace{15mm} {\rm for}\hspace{8mm} n=-2,
 \end{array}\right.
\end{equation}
where $\phi_i$ is an integration constant and
\begin{equation}\label{V0-caseI}
V_0\equiv2\beta D(1-D)h^2(D).
\end{equation}
However, it has been shown that the following relations are independent of $n$:
\begin{eqnarray}\label{new74}
\rho\!\!&=&\!\!\left(1-D\right)^2h^2(D)
\left[-Dh(D)(t-t_i)+1\right],\\\nonumber\\
 p\!\!&=&\!\!\left(1-D\right)h^2(D)\times\left[D\left(1-D\right)h(D)(t-t_i)-1\right],
\label{new75}
\\\nonumber\\
\rho_{\phi}\!\!&=&\!\!\frac{D(1-D)h^2(D)}{2}\times\left[1+2\left(1-D\right)h(D)(t-t_i)\right],
\label{ro-phi-deSit}
\\\nonumber\\
p_{\phi}\!\!&=&\!\!\frac{D(1-D)h^2(D)}{2}\times\left[1-2\left(1-D\right)h(D)(t-t_i)\right].
\label{p-phi-deSit}
\end{eqnarray}
Hence substituting the components of the induced matter and the matter associated
with the SB scalar field into equation \eqref{ops5}, the GD equation of the
null vector fields past directed, in the context of the MSBT for $f(D)=0$, is given by
\begin{eqnarray}\label{GDE-MSBT}
\frac{d^2\eta}{dz^2}+\frac{2}{1+z}\frac{d\eta}{dz}=0, \hspace{10mm} \forall n.
\end{eqnarray}
%The
Equation \eqref{GDE-MSBT} yields an exact solution as
\begin{eqnarray}\label{GDE-MSBT-sol}
\eta(z)=-\frac{C_1}{1+z}+C_2,\hspace{10mm} \forall n,
\end{eqnarray}
where $C_1$ and $C_2$ are integration constants, which have the same units of $\eta$.
Therefore, the observer area distance $r_0(z)$ associated with this case is given by
\begin{eqnarray}\label{mattig-exp}
r_0(z)=\frac{z}{H_0(1+z)}, \hspace{10mm} \forall n.
\end{eqnarray}

\subsubsection{GD equation for Power-law solution}
For the case where $f(D)\neq0$, the scale factor $a(t)$ is given by a
power-law form in terms of the cosmic time. Concretely, from
equations (\ref{new18-2}), the SB scalar field is obtained \cite{RPSM20}:
\begin{equation}\label{new20-22}
 \phi(t)=\left \{
 \begin{array}{c}
\left\{\frac{n+2}{2f(D)}{\rm ln}\left[\tilde{h}(D)(t-t_i)\right]\right\}^{\frac{2}{n+2}}
 \hspace{9mm} {\rm for}\hspace{8mm} n\neq-2,\\\\
  \left[\tilde{h}(D)(t-t_i)\right]^{\frac{1}{f(D)}}
  \hspace{21mm} {\rm for}\hspace{8mm} n=-2,
   \end{array}\right.
\end{equation}
where
\begin{equation}\label{h-tild}
\tilde{h}(D) \equiv \frac{c_1f(D)}{a_i^{D-1}\psi_i}.
\end{equation}
%Moreover,
The scale factor $a$ and the scalar field $\psi$ are obtained in terms of the cosmic time $t$:
\begin{eqnarray}\label{new29}
 a(t)\!\!&=\!\!&a_i \left[\tilde{h}(D)(t-t_i)\right]^{r},\hspace{8mm} \forall n,\\\nonumber
 \\\label{new29-2}
   \psi(t)\!\!&=&\!\! \psi_i \left[\tilde{h}(D)(t-t_i)\right]^{m},\hspace{8mm} \forall n.
 \end{eqnarray}
In equations \eqref{new29} and \eqref{new29-2}, $r$ and $m$ were defined as
 \begin{eqnarray}\label{r-m}
r\equiv\frac{\gamma}{f(D)},\hspace{3mm} m\equiv\frac{\beta}{f(D)},\hspace{3mm} {\rm where} \hspace{3mm} m+(D-1)r=1.
 \end{eqnarray}

 The induced potential is given by \cite{RPSM20}
 \begin{equation}\label{new32}
V(\phi)=\left \{
 \begin{array}{c}
-\frac{\tilde{V}_0}{2f(D)}\,{\rm Exp}\left[\frac{-4f(D)}{n+2}\,\phi^{\frac{n+2}{2}}\right]
\hspace{7mm} {\rm for}\hspace{5mm} n\neq-2,\\\\
 -\frac{\tilde{V}_0}{2f(D)}\phi^{-2f(D)}
\hspace{24mm} {\rm for}\hspace{5mm} n=-2,
 \end{array}\right.
\end{equation}
where $\tilde{V}_0$ is related to the other parameters of the model as
\begin{equation}\label{V0}
\tilde{V}_0\equiv 2c_1^2\beta{\cal W}a_i^{2(1-D)}\psi_i^{-2}.
 \end{equation}
One can show that
\begin{eqnarray}\label{new46}
\rho&=&-\frac{D(D-1)m r^2}{2 (t-t_i)^{2}},
\hspace{4mm} p=-\frac{D m r\left(1+m\right)}{2 (t-t_i)^{2}}, \\\nonumber\\
\label{ro-phi}
\rho_{\phi}&=&\frac{\left[(D-1)r\right]^2
\left[2m+(D-2)r\right]}{2(t-t_i)^2},\\\nonumber\\
p_{\phi}\!\!\!&=&\!\!\frac{\left[(D-1)r\right]
\left[2m+(D-1)r\right]\left[2m+(D-2)r\right]}{2(t-t_i)^2},
\label{p-phi}
\end{eqnarray}
which are valid for all values of $n$.

Substituting $\rho$, $p$, $\rho_{\phi}$ and $p_{\phi}$ from relations
 \eqref{new46}-\eqref{p-phi} into \eqref{ops5}, the GD equation for this
 case will be exactly the differential equation \eqref{GDE-power} with an exact solution  \eqref{GDE-power-sol}.
Using \eqref{mattig-1} for \eqref{GDE-power-sol},
one can show that the observer area distance is given by \eqref{GDE-power-r}.
 However, for this case, it is important to note that $Q$ is given by
%\begin{widetext}
\begin{eqnarray}\label{Q}
{\cal Q}= {\cal Q}(\beta,{\cal W},D)\equiv \frac{(D-1)}{{\cal W}}
\left[{\cal W}+\beta^2\pm\beta\sqrt{\beta^2
+\left(\frac{D-2}{D-1}\right){\cal W}}\,\,\right]=\frac{1-m}{(D-1)}.
\end{eqnarray}
%\end{widetext}
 It is worth to depict the behavior of the deviation vector as well as
 the observer area distance. For this aim, let us employ the following procedure.

 In \cite{RPSM20}, it has been shown that for specific allowed
 ranges of the independent parameters of the model, i.e., either $\{\beta>0,
 2(D-1)\beta\gamma<{\cal W}<(D-1)\beta\gamma<0\}$
 or $\{\beta<0,\,\, 2(D-1)\beta\gamma<{\cal W}<(D-1)\beta\gamma<0\}$,
 it is feasible to obtain an accelerating scale factor which could be
 applicable for the present universe.

\begin{figure}
\centering\includegraphics[width=2.6in]{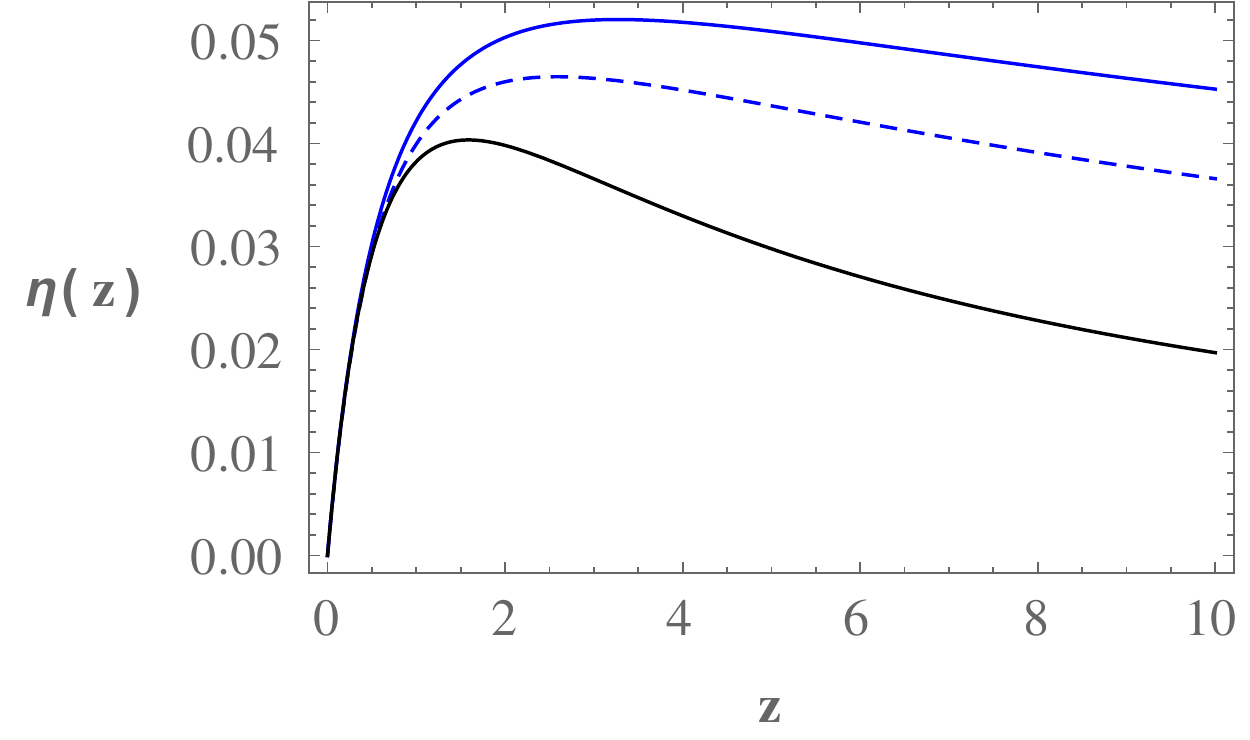}
\centering\includegraphics[width=2.6in]{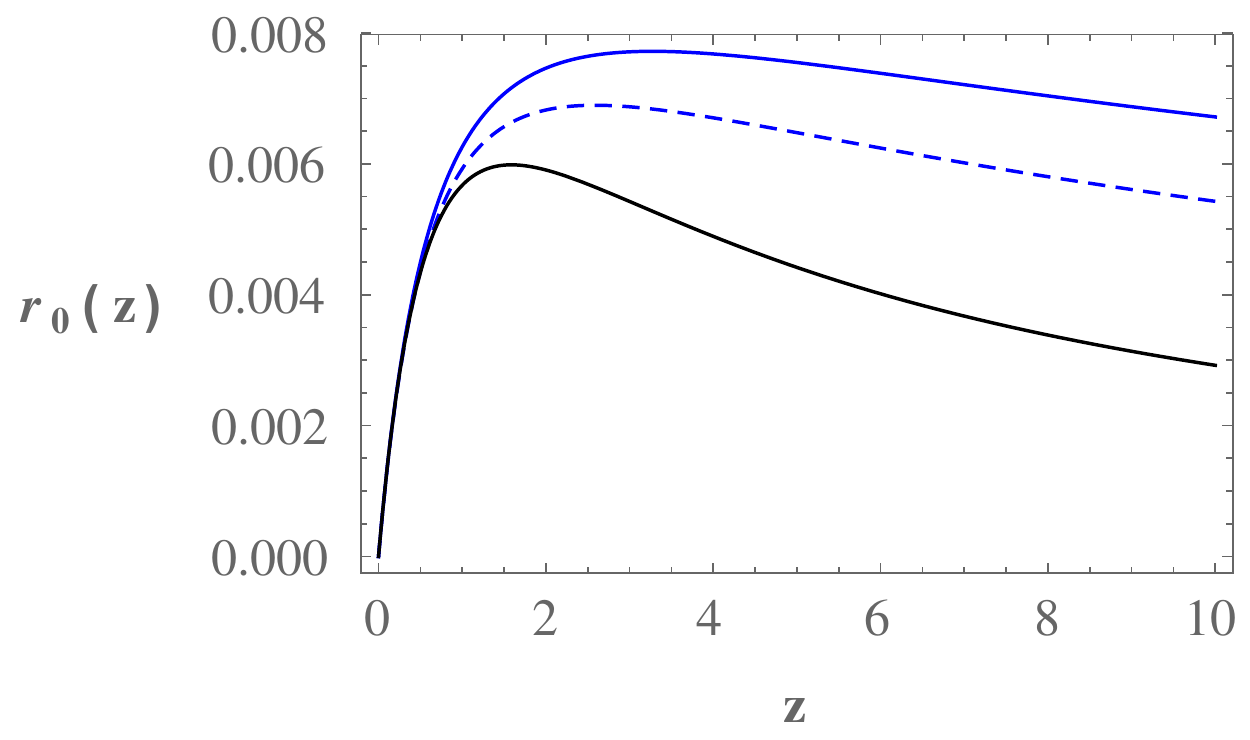}
\caption{{\footnotesize The behavior $\eta(z)$ (the left panel)
and $r_0(z)$ (the right panel) associated with the null vector fields
with FLRW background in the context of $\Lambda$CDM model (the black curves) and the MSBT in four
dimensions (the dashed and solid blue curves) with $f(D=4)\neq 0$. We use units of $H_0^{-1}$, $8 \pi G=1$, and %we have
assume $\eta(0)=0$, $d\eta(z)/dz\mid_{_{z=0}}=0.1$, $H_0=67.4 KM/s/Mps$
and $q_0=-0.55$ (the solid blue curves) and $q_0=-0.4$ (the dashed curves).
}}
%\foreignlanguage{english}
{\label{eta-r-MSBT-power}}
\end{figure}

 Let us express
 ${\cal Q}$ in terms of the deceleration parameter (which reads for our herein power-law solution as $q=1/r-1$): %from
  using equations \eqref{r-m} and \eqref{GDE-power-sol}, we obtain ${\cal Q}=q+1$.
   Moreover, we would use the same initial conditions used before, i.e., $\eta(0)=0$
  and $d\eta(z)/dz\mid_{z=0}=0.1$, which leads to \eqref{GDE-power-IC}.
We restrict our attention to the four-dimensional case for
which we can use the recent observational data reported in \cite{PL.CosPar18}.
 In figure \ref{eta-r-MSBT-power}, we
 plot the behavior of $\eta(z)$ and $r_0(z)$ for this
 case and compare them with those of the $\Lambda$CDM model.

\section{Discussion and Conclusions}
\label{Concl}

In this paper, we computed and  investigated the
general form for the
GD equation in the framework of (i) an extended version of
the conventional SB theory \cite{SB85l} (i.e., not only
 we have considered a general scalar potential, but also
 assumed an  arbitrary number of spatial dimensions) and
%(ii) the $\Lambda$CDM model \bl{as well as a phantom scenario}, and
(ii) the MSBT theory \cite{RPSM20}.
%By introducing a
%generalized version of the standard SB theory, we
%have firstly obtained the general form for the GD equation.
We have employed  the ordinary matter as a perfect fluid and chosen
the line-element such  that the  Weyl tensor vanishes.
%Assuming the spacetime geometry is described by
%a spatially flat FLRW metric, the resulted formalism yields the
%generalized version of the Pirani equation.
We  focused on two particular
case studies: fundamental observers
%(for which the Raychaudhuri equation is obtained in $D$ dimensions)
  and null vector fields.
  Subsequently, for the particular case where the SB scalar field takes constant values,
 the GD equation of the null vector fields reduces to the
corresponding one in the $\Lambda$CDM model, as expected.

To apply the GD equation for the simplest case study,
%that we retrieved in either of the SB theory we have mentioned,
we assumed that the SB scalar field dominates
the dynamics.
% (cf subsection \ref{new-sol}).
 In this regard,
we have extracted two cosmological
 scenarios, in the form of {\it new} exact solutions.
 We have shown that, in a particular
 case, these models reduce to those retrieved in the
 context of GR where a scalar field is minimally coupled to gravity.
 In the particular case where $n=0$, ${\cal W}=1$ and $D=4$
 (or using any other equivalent conditions, which can produce such a case), the Lucchin-Mataresse and
  Barrow--Burd--Lancaster--Madsen models are recovered from
  solutions I and II, respectively (please, see subsection \ref{new-sol}).
  %Moreover, in subsection \ref{GR}
  We have also used
a specific form of a perfect fluid
which is described by non-interacting dust and radiation.

%In  another application concerning Section \ref{SetUp},
 We study a phantom dark energy model
  in the context of an SB theory. Assuming this setting
 is applicable
  for the late time accelerating universe, we used
  the GD equation for small values of the redshift parameter.
   We have employed the energy
  conditions and %the
  recent observational data to find the allowed values
  of the corresponding parameters of the model.
  Such a procedure assisted us to depict the evolution of the
  deviation vector as well as the observer
  area distance against the redshift parameter.
  As it is well-known that phantom dark energy
  models are able to alleviate the $H_0$ tension, we have therefore considered two sets
  of the observational data, which have been reported by the Planck
collaboration and by the SH0ES collaboration, to plot the behavior of $\eta(z)$ and $r_0(z)$.
   We have also compared \textbf{their} behavior
  %$\eta(z)$ and $r_0(z)$ retrieved   for  that phantom dark energy model
  with those \textbf{plotted} according to the $\Lambda$CDM model.
  Our endeavors have shown that the general behavior of these quantities are similar for all models.

%One of %the
%our main objectives %of this investigation
 We also studied the GD equation in the MSBT \cite{RPSM20}.
In this respect, %after providing a brief review of this framework,
we have shown that all the formalism
%obtained in section \ref{SetUp}
 in the context of the generalized SB theory could also
be applied for the MSBT.
%This %is a natural result
%was expected since the standard SB
%theory (including a scalar potential) and the MSBT are obtained from a sole action.
%The only difference is that the EMT as well as the potential of the MSBT
%emerge strictly from the geometry rather than {\it ad hoc} assumptions.
%As another application of the GDE obtained in section \ref{SetUp},
Subsequently, we have retrieved the corresponding cosmological exact solutions
within the MSBT framework, namely within a spatially flat FLRW background. We have
investigated the GD equation for a null vector field past directed, specifically for
those mentioned cosmological solutions,
 and plotted the behavior of $\eta(z)$ and $r_0(z)$.

We have shown that the behaviour of
the plotted observables (i.e., $\eta(z)$ and $r_0(z)$ for small values of the redshift parameter),
either appraising them quantitative or qualitative
associated with the new cosmological solutions extracted in
the generalized SB theory and the MSBT (which have been also contrasted
with either $\Lambda$CDM model or a phantom dark
energy model), are all similar. However, it is
important to note that only the latter could be considered as  fundamental.
 In contrast to the other cosmological
settings investigated in this paper, the EMT as well as the scalar potential present in the MSBT
are not added by {\it ad hoc} assumptions to the action, but instead, they emerge strictly from dimensional reduction
from the geometry, including the extra spatial dimensions %Such real quantities
%(although, contrary to the conventional theories, the reduction procedure
%dictates particular forms for them) could provide an
%appropriate description for the late time universe
\cite{RPSM20}.
Let us emphasize that the analysis associated with the GD equation in  the MSBT
%(specially, those of the case study in this work,
i.e., the null vector
fields appraisal,  is fully  consistent with the current observational data.

Let us close this section with the following comments.
\begin{itemize}

\item  We should note that, for the sake of generality, all of our calculations have
 been done in arbitrary dimensions. Although, as a toy model, it is easy to plot the figures
 for any values of $D\geq3$, using the observational data we have examined our herein model
 only for the cases with $D=4$.
  \item
For a general case, it may not possible to consider transformations by which the
action \eqref{induced-action} proceeds to a corresponding case with a canonical kinetic term.
Notwithstanding, because of the importance of this
point, let us assume a particular case such that the coupling function
only takes positive values, i.e., ${\cal W}\phi^n\equiv J(\phi)>0$.
In this case, defining a canonical scalar
field as $d\tilde{\phi}= \sqrt{J(\phi)}d\phi$,
the gravitational sector of the SB model \eqref{induced-action} becomes \cite{YZ21}
\begin{eqnarray}
 {\cal S}^{^{(D)}}=\int d^{^{\,D}}\!x \sqrt{-g}\,
 \Big[R^{^{(D)}}-g^{\alpha\beta}\,({\nabla}_\alpha\tilde{\phi})
 ({\nabla}_\beta\tilde{\phi})-U(\tilde{\phi})\Big]
\label{canon-1}.
\end{eqnarray}
Note that the canonical potential
and $V(\phi)$ are related as $U[\tilde{\phi}(\phi)]=V(\phi)$.
It is straightforward to show that the coupling function $J(\phi)$ can
be expressed in terms of the potentials, such that the action \eqref{canon-1} is rewritten as
\begin{eqnarray}
 {\cal S}^{^{(D)}}=\int d^{^{\,D}}\!x \sqrt{-g}\,
 \Big[R^{^{(D)}}-\left(V_{,\phi}\frac{dU^{-1}(V(\phi))}{dV(\phi)}\right)^2g^{\alpha\beta}\,({\nabla}_\alpha\phi)
 ({\nabla}_\beta\phi)-V({\phi})\Big],
\label{canon-2}
\end{eqnarray}
where $U^{-1}$ is the inverse function of $U$.
It should also be noted that actions \eqref{canon-1} and \eqref{canon-2} are equivalent and
they determine the same predictions \cite{YZ21}.
However, it seems that for any non-canonical model with specified coupling function
(see e.g., the SB model with $J(\phi)= {\cal W}\phi^n>0$), it is important to note that, using the above transformation
for getting the canonical kinetic term, restricted us to take a special canonical potential, see \eqref{canon-2}.
From what we pointed out above, we find that our discussions associated with the GD equation in
the SB context, in particular cases, can also be applied for the
gravitational models whose actions possess a canonical kinetic term.
We emphasize that, to the best of our knowledge, the GD
equation associated with the latter case has not yet  been investigated.

\item Furthermore, we should note that it is not easy to find transformations by which the
field equations of our generalized SB theory (for
general values of ${\cal W}$ and $n$) can transform to the corresponding ones of the
Brans-Dicke theory. However, %it is seen that
the GD equations \eqref{GDE-gen-2}, \eqref{fun-3} and
\eqref{ops5} bear close resemblance to those obtained
in the context of the Brans-Dicke theory.
More concretely, letting
\begin{eqnarray}
\rho_{_{\rm eff}}\rightarrow \frac{1}{\varphi}\left(\rho+\rho_{\varphi}\right),
\hspace{10mm}p_{_{\rm eff}}\rightarrow \frac{1}{\varphi}\left(p+p_{\varphi}\right),
\label{SB-BD-1}
\end{eqnarray}
where $\rho_{\varphi}$ and $\rho_{\varphi}$ stand for the energy
density and pressure associated with the BD scalar field $\varphi$, respectively:
\begin{eqnarray}
\label{rho-phi-BD}
\rho_\varphi\!\!&\equiv\!\!&\frac{\omega}{2}
\frac{\dot{\varphi}^2}{\varphi}+\frac{V(\varphi)}{2}-(D-1)H\dot{\varphi},
\\\nonumber\\
p_\phi\!\!&\equiv\!\!&\frac{\omega}{2}
\frac{\dot{\varphi}^2}{\varphi}-\frac{V(\varphi)}{2}+\ddot{\varphi}+(D-2)H\dot{\varphi},
\label{p-phi-BD}
\end{eqnarray}
(where $\omega$ is the BD coupling parameter),
then equations \eqref{GDE-gen-2}, \eqref{fun-3} and \eqref{ops5} transform
to the corresponding ones obtained in the context of
the BD theory, for more details we refer the reader to \cite{RS21}.

\item
It is important to note that equation \eqref{phan-GDE-2} is valid (as the GD equation
associated with the past directed null vector field corresponding
to the spatially flat FLRW metric) not only
for the phantom dark energy model, but also for any cosmological
model investigated in the context of the generalized
SB theory in arbitrary dimensions.
 Such a significant consequence can
be easily shown from using equations \eqref{fun-Ray}, \eqref{fun-Fri-1},
\eqref{cons}, \eqref{cons-phi}, \eqref{ops5} and the definition of the deceleration
parameter. However, we should emphasize that
the $q(z)$ is, obviously, a model dependent quantity.

\item
One of the biggest shortcomings of GR is predicting existence of
singularities, which can be indicated by singularity theorems, see, for instance, \cite{P65,HP70}.
The Raychaudhuri equation has been employed as one of the
important ingredients to prove such theorems.
%``{\it If the strong energy condition (SEC) holds, an initially converging
%timelike geodesic congruence focuses within finite affine parameter value \cite{}}'' ;see also [5,6].
%This statement, in turn, follows from the Raychaudhuri equation.
A congruence singularity, whether or not could be considered as a curvature
singularity, is caused by focusing of congruence, by which, together
with a few additional reasonable conditions
on a spacetime, the singularities emerge.
In GR, the {\it convergence condition} $R_{\mu\nu}u^\mu u^\nu\geq 0$ (which leads to
geodesic focusing from an attractive gravity) is retrieved from the SEC.
As the field equations associated with alternative theories to GR are
different, therefore, even if the SEC is satisfied, it is possible
that the convergence condition is violated \cite{BMDHU18,CDB21}.
Let us focus on our herein model.
For the case established in part (iii) of
Section \ref{SetUp}, from using equation \eqref{gen-Ricc-tensor}, we obtain
\begin{eqnarray}\label{CC-1}
R_{\mu\nu}u^\mu u^\nu=\left[T_{\mu\nu}-\left(\frac{T}{D-2}\right)g_{\mu\nu}+
{\cal W}\phi^n(\nabla_\mu\phi)(\nabla_\nu\phi)
+\left(\frac{1}{D-2}\right)g_{\mu\nu}V(\phi)\right]u^\mu u^\nu.
\end{eqnarray}
For the case of perfect fluid (which was discussed in
part (v) of Section \ref{SetUp}), equation \eqref{CC-1} for the geodesic
congruences with timelike and null vector fields reduces to
\begin{equation}\label{CC-2}
 R_{\mu\nu}u^\mu u^\nu=\left \{
 \begin{array}{c}
\frac{1}{D-2}\left[(D-3)\rho+(D-1)p\right]+{\cal W}\phi^n \dot{\phi}^2-\frac{V(\phi)}{D-2}
\hspace{8mm} {\rm (timelike) },\\\\
  \frac{1}{D-2}\left(\rho+p+{\cal W}\phi^n \dot{\phi}^2\right)
  \hspace{31mm}  {\rm (null)},
   \end{array}\right.
\end{equation}
which can also be read from equations \eqref{fun-3} and \eqref{BD-null}.
Obviously, without considering a specific exact solution, we cannot proceed discussion.
In this regard, it is straightforward to
determine the overall signature of $R_{\mu\nu}u^\mu u^\nu$ for
our exact solutions obtained in Sections \ref{SB-GDE} and \ref{MSBT}.
Such an investigation to study the violation of convergence
condition may constrain the parameters of the model.

\item
 Finally, it is worth noting that further investigation is
 required to obtain concrete constraints on the SB coupling
 parameter ${\cal W}$, so  to be consistent with current observational data.
 Such a procedure is not in the scope of this paper
 and might be presented in our future investigations.

\end{itemize}

\section{ACKNOWLEDGMENTS}
We would like to thank the anonymous referee
for valuable comments, which have led to
improve the manuscript.
PVM and SMMR acknowledge the FCT grants UID-B-MAT/00212/2020 and UID-P-MAT/00212/2020 at CMA-UBI plus the COST Action CA18108 (Quantum gravity phenomenology in the multi-messenger approach).
The work of M.S. is supported in part by the Science and Technology Facility Council
(STFC), United Kingdom, under the research grant ST/P000258/1.
%%%%%%%%%%%%%%%%%%%%%%%%%%%%%%%%%%%%%%%%%%%%%%%%%%%%%%%%%%%%%%%%%%%%%

\end{document}